\documentclass[10pt]{article}

\usepackage{ragged2e}
\usepackage{geometry}
\makeatletter
\renewcommand\normalsize{%
   \@setfontsize\normalsize\@xpt{14}%
   \abovedisplayskip 10\p@ \@plus2\p@ \@minus5\p@
   \abovedisplayshortskip \z@ \@plus3\p@
   \belowdisplayshortskip 6\p@ \@plus3\p@ \@minus3\p@
   \belowdisplayskip \abovedisplayskip
   \let\@listi\@listI}
\normalsize
\renewcommand\small{%
   \@setfontsize\small\@ixpt{12}%
   \abovedisplayskip 8.5\p@ \@plus3\p@ \@minus4\p@
   \abovedisplayshortskip \z@ \@plus2\p@
   \belowdisplayshortskip 4\p@ \@plus2\p@ \@minus2\p@
   \def\@listi{\leftmargin\leftmargini
               \topsep 4\p@ \@plus2\p@ \@minus2\p@
               \parsep 2\p@ \@plus\p@ \@minus\p@
               \itemsep \parsep}%
   \belowdisplayskip \abovedisplayskip
}
\renewcommand\footnotesize{%
   \@setfontsize\footnotesize\@viiipt{10}%
   \abovedisplayskip 6\p@ \@plus2\p@ \@minus4\p@
   \abovedisplayshortskip \z@ \@plus\p@
   \belowdisplayshortskip 3\p@ \@plus\p@ \@minus2\p@
   \def\@listi{\leftmargin\leftmargini
               \topsep 3\p@ \@plus\p@ \@minus\p@
               \parsep 2\p@ \@plus\p@ \@minus\p@
               \itemsep \parsep}%
   \belowdisplayskip \abovedisplayskip
}
\renewcommand\scriptsize{\@setfontsize\scriptsize\@viipt\@viiipt}
\renewcommand\tiny{\@setfontsize\tiny\@vpt\@vipt}
\renewcommand\large{\@setfontsize\large\@xipt{15}}
\renewcommand\Large{\@setfontsize\Large\@xiipt{16}}
\renewcommand\LARGE{\@setfontsize\LARGE\@xivpt{18}}
\renewcommand\huge{\@setfontsize\huge\@xxpt{30}}
\renewcommand\Huge{\@setfontsize\Huge{24}{36}}
\makeatother

\usepackage[tableposition=top]{caption}
\usepackage{ifpdf}
\usepackage{xcolor}
\usepackage{paralist}
\usepackage{tabularx}
\usepackage{booktabs}
\usepackage{multirow}
\usepackage{graphicx}
\usepackage{sidecap}
\usepackage{natbib}
\renewcommand{\cite}{\citep}
\usepackage{amssymb}
\usepackage{amsmath}
\usepackage{lscape}
\makeatletter
\def\vcdots{\vbox{\baselineskip4\p@ \lineskiplimit\z@
    \kern3\p@\hbox{.}\hbox{.}\hbox{.}\kern3\p@}}
\makeatother
\usepackage{caption}
\usepackage{array}
\usepackage{url}
\usepackage{float}

\ifpdf
  \usepackage{epstopdf}
	\usepackage{microtype}
\fi
\graphicspath{{./figures/}}

\usepackage{algorithm}
\usepackage[noend]{algorithmic}

\colorlet{TufteRed}{red!80!black}

\definecolor{theblue}{RGB}{0,0,180}
\colorlet{thered}{TufteRed}

\usepackage{dgleich-setup}
\usepackage[notheorems]{dgleich-math}

\renewcommand{\mat}{\boldsymbol}

\usepackage{mathtools}
\usepackage[amsmath,thmmarks]{ntheorem}

\usepackage{etoolbox}
\apptocmd{\sloppy}{\hbadness 10000\relax}{}{}
\apptocmd{\thebibliography}{\raggedright}{}{}

\theoremstyle{break}
\theoremheaderfont{\normalfont\bfseries\itshape}
\theoremprework{\vspace{7pt}\list{}{\rightmargin\leftmargin}%
   \item\relax}
\theorempostwork{\endlist\vspace{7pt}}
\theorembodyfont{\normalfont}
\theoremseparator{}

\newtheorem{theorem}{Theorem}

\theoremclass{theorem}

\theoremclass{theorem}

\theoremclass{theorem}

\theoremclass{theorem}
\newtheorem{corollary}[theorem]{Corollary}
\theoremclass{theorem}
\newtheorem{proposition}[theorem]{Proposition}

\theoremclass{theorem}
\newtheorem{definition}[theorem]{Definition}

\theoremstyle{nonumberplain}
\theoremheaderfont{\normalfont\itshape}
\theorembodyfont{\upshape}
\theoremseparator{\hspace{1ex}}
\theoremsymbol{\rule{1ex}{1ex}}
\theoremprework{\vspace{-7pt}}
\theoremclass{proof}

\theoremclass{proof}
\theoremstyle{plain}
\theoremheaderfont{\normalfont\bfseries\sscshape}
\theorembodyfont{\upshape}
\theoremseparator{\hspace{1ex}}
\theoremsymbol{\ensuremath{\filleddiamond}}

\ifpdf%
  \RequirePackage[colorlinks,pdfdisplaydoctitle
      ,citecolor=theblue
      ,linkcolor=theblue
      ,urlcolor=theblue
      ,hyperfootnotes=false,pdftex,pagebackref]{hyperref}%
\else
  \RequirePackage[colorlinks,pdfdisplaydoctitle
      ,citecolor=theblue
      ,linkcolor=theblue
      ,urlcolor=theblue
      ,hyperfootnotes=false,dvipdfm,pagebackref]{hyperref}
\fi
\renewcommand*{\backref}[1]{}
\renewcommand*{\backrefalt}[4]{%
  \ifcase #1 %
    No citations.%
  \or
    Cited on page #2.%
  \else
    Cited on pages #2.%
  \fi
}

\newcommand{\secref}[1]{\hyperref[sec:#1]{section~\ref*{sec:#1}}}
\newcommand{\Secref}[1]{\hyperref[sec:#1]{Section~\ref*{sec:#1}}}
\newcommand{\secrefp}[1]{\hyperref[sec:#1]{(section~\ref*{sec:#1})}}

\newcommand{\thmref}[1]{\hyperref[thm:#1]{theorem~\ref*{thm:#1}}}
\newcommand{\Thmref}[1]{\hyperref[thm:#1]{Theorem~\ref*{thm:#1}}}
\newcommand{\thmrefp}[1]{\hyperref[thm:#1]{(theorem~\ref*{thm:#1})}}

\newcommand{\figrefp}[1]{\hyperref[fig:#1]{(figure~\ref*{fig:#1})}}
\newcommand{\figref}[1]{\hyperref[fig:#1]{figure~\ref*{fig:#1}}}
\newcommand{\Figref}[1]{\hyperref[fig:#1]{Figure~\ref*{fig:#1}}}
\usepackage{multicol}
\usepackage{tabularx}
\usepackage{graphicx}
\usepackage{subcaption}

\usepackage{enumitem}

\newcommand{\vecn}[1]{\ensuremath{\textbf{#1}}}
\newcommand{\vvk}[2]{ \vecn{#1}^{(#2)} }
\newcommand{\hvv}[1]{\hat{\vecn{#1}}}
\newcommand{\inv}{^{-1}}
\newcommand{\floor}[1]{\lfloor #1 \rfloor }

\newcommand{\ceil}[1]{\lceil #1 \rceil}

\renewcommand{\abstract}{}

\title{Localization in Seeded PageRank
}
\author{David F.~Gleich, Kyle Kloster, Huda Nassar \\
	\{Computer Science, Mathematics, Computer Science\} Department \\
	Purdue University\\
	\{dgleich, kkloste, hnassar\}@purdue.edu }

\begin{document}

\maketitle

\begin{abstract} \small
Seeded PageRank is an important network analysis tool for identifying and studying regions nearby a given set of nodes, which are called seeds. The seeded PageRank vector is the stationary distribution of a random walk that randomly resets at the seed nodes. Intuitively, this vector is concentrated nearby the given seeds, but is mathematically non-zero for all nodes in a connected graph.  We study this concentration, or localization, and show a sublinear upper bound on the number of entries required to approximate seeded PageRank on all graphs with a natural type of skewed-degree sequence---similar to those that arise in many real-world networks.
Experiments with both real-world and synthetic graphs give further evidence to the idea that the degree sequence of a graph has a major influence on the localization behavior of seeded PageRank. Moreover, we establish that this localization is non-trivial by showing that complete-bipartite graphs produce seeded PageRank vectors that cannot be approximated with a sublinear number of non-zeros. 
\end{abstract}

\section{Introduction}\label{sec:intro}

The PageRank vector of a graph has a number of different derivations, but the most famous is via the random surfer process~\cite{page1999-pagerank}.
As PageRank's random surfer moves between the nodes of a graph, at each node with probability $\alpha$ the surfer chooses to follow a random neighbor (chosen uniformly at random over all neighbors), and otherwise it resets its state with probability $1-\alpha$.
On a reset, the surfer \emph{teleports} to a new state chosen from a \emph{reset distribution}. The PageRank vector is the long-term fraction of time the process spends at each node, or, equivalently, the stationary distribution. (The PageRank process is a special case where these two concepts are the same.)  \emph{Seeded} PageRank refers to PageRank vectors where the reset distribution is a single node. Originally, PageRank vectors with these simple reset distributions were called personalized PageRank vectors due to PageRank's origins on the web, but we feel ``seeded PageRank'' is more appropriate given the vast number of uses of PageRank outside the web~\cite{Gleich-2015-prbeyond}. Furthermore, large entries in seeded PageRank vectors are often used to identify \emph{regions} nearby the seed node that support various network analysis tasks, see the survey~\cite{Gleich-2015-prbeyond} for many examples.

A seeded PageRank vector is determined by three inputs:
\begin{compactenum}
\item the underlying graph, represented by a column-stochastic probability transition matrix $\mP$ where $P_{j,i} = 1/\text{degree}(i)$ for each directed edge $(i,j)$ between nodes $i$ and $j$ (we label all nodes with a unique integer index and assume that all nodes have degree at least 1);
\item a parameter $\alpha$ that determines the reset probability $(1-\alpha)$; and
\item a seed node, $s$.
\end{compactenum}
In our notation, the vector $\ve_s$ is the vector of all zeros with a single 1 in the
position corresponding to node $s$. The seeded PageRank vector
$\vx$ is then the solution of the linear system:
\[ (\mI - \alpha \mP) \vx = (1-\alpha) \ve_s. \]
When the network is strongly connected, the solution $\vx$ is non-zero for all nodes. This is a consequence of the fact that there is a non-zero probability of walking from the seed to any other node in a strongly connected network.

\begin{figure}
\includegraphics[width=0.5\linewidth]{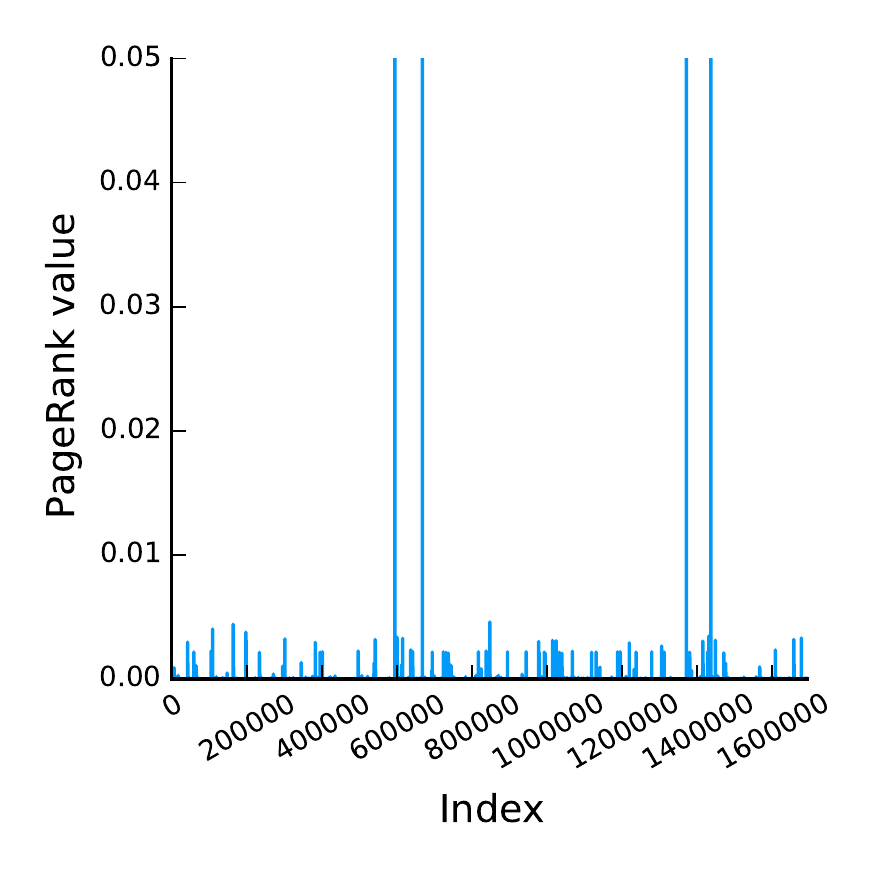}%
\includegraphics[width=0.5\linewidth]{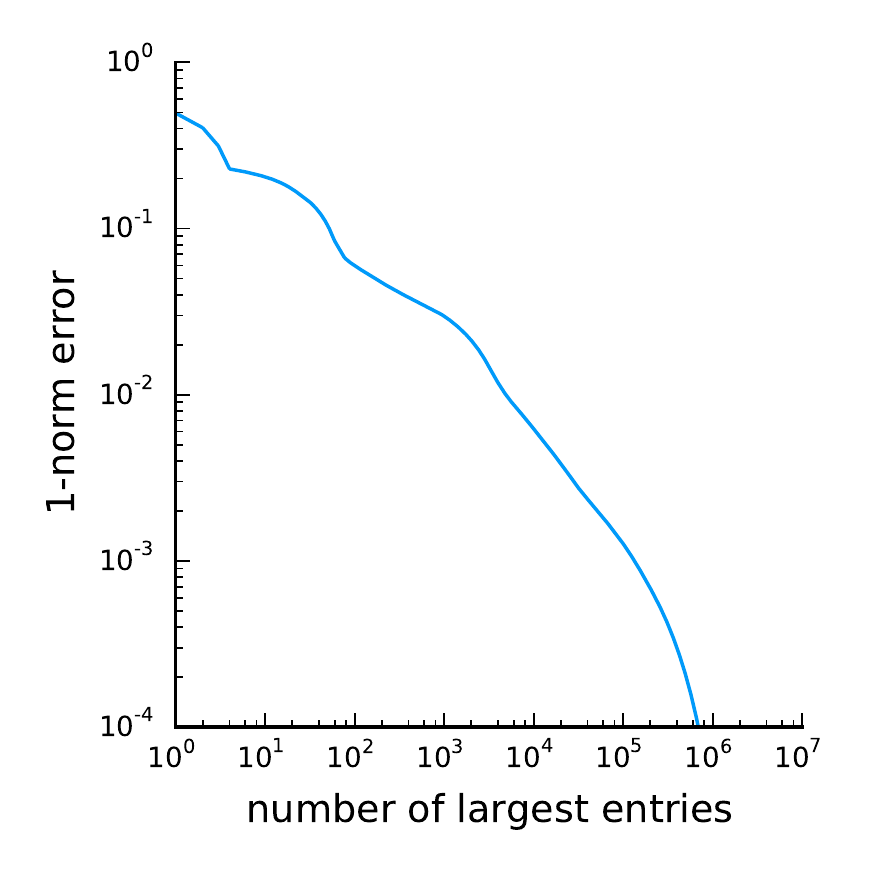}
\caption{At left, we display a seeded PageRank vector with $\alpha=0.5$ from the as-skitter graph based on \emph{index order} (the arbitrary numerical labels of each node in the graph). The peaks indicate that there are only four extremely large entries. At right, we show the same vector in terms of the \emph{accuracy} that results from using the largest $k$ values in an approximation. This plot shows that we can get a $10^{-2}$, globally accurate, approximation of the vector with only about 10,000 entries out of 1.7 million. Our goal in this paper is to understand this type of localization in PageRank vectors. }
\label{fig:example}
\end{figure}

A vector or sequence is localized when we can globally approximate all of the entries using an asymptotically vanishing subset. For example, we can globally approximate a convergent infinite geometric series with only a few of the largest entries.
We seek a theoretical understanding of localization in seeded PageRank vectors and are concerned with the minimal number of entries required to approximate these vectors. As a concrete example of localization, in Figure~\ref{fig:example} we plot the seeded PageRank vector for a randomly chosen node in a 1.7 million node graph with 22 million edges. (The graph is the largest connected component of the as-skitter network~\cite{snap,leskovec2005shrinking-diameter}.)
The true PageRank solution is non-zero everywhere, and the two plots illustrate different aspects of what localization looks like in a ``large vector.''
In this case, the plots show that we can approximate the seeded PageRank vector $\vx$ with 1.7 million entries using only roughly 10,000 of the entries and attain a global accuracy of $10^{-2}$. (Formally, $\normof[1]{\vx - \vy} \le 10^{-2}$ where $\vy$ is composed of just the largest 10,000 entries from $\vx$.)
Our goal is to understand when this behavior is possible.

Our results establish that all graphs with a specific, but realistic, type of degree distribution have seeded PageRank vectors that can be approximated by vectors with only a sublinear number of non-zero entries. At this point, we state a simplified version of our main theorem to give the reader an intuition for how the bound behaves. The setting, which we describe more precisely in the full theorem (Theorem~\ref{thm:non-zeros}), is an $n$ node graph with max degree $d$ and what we define as a rank-skewed degree sequence (Definition~\ref{def:degree-sequence}). To approximate a seeded PageRank vector with parameter $\alpha$ and accuracy $\eps$, it suffices to use no more nonzeros than
\begin{equation}
\tilde{O}( d^{1/p} (\tfrac{\alpha^2}{\eps})^{1/(1-\alpha)}),
\end{equation}
where $d$ is the maximum degree, $p$ reflects a parameter of degree sequence that is empirically around $0.5$ to $1$, and the $\tilde{O}$ notation hides logarithmic factors.
Importantly, the only quantity in this bound that may depends on $n$ is the max degree, $d$. This means that the bound scales sublinearly with $n$ as long as the $d^{1/p}$ is sublinear.
Our proof is constructive and yields an algorithm to identify this sublinear-sized set. Thus, these results show that we can \emph{locally compute seeded PageRank vectors for a realistic class of graphs}.


Our code and data are available for others to reproduce our empirical findings:
\begin{center}
\url{https://github.com/nassarhuda/pprlocal}
\end{center}

\subsection{Differences from the WAW2015 Conference Paper}
The results in this paper are closely related to our prior WAW2015 conference paper~\cite{nassar2015strong}, where we showed the first localization result for PageRank vectors beyond the constant-degree case~\cite{bar2008local}.  The differences in this paper are relatively minor in that we improve the localization bounds in small, meaningful ways. Furthermore, we have stronger characterizations of the space where localization is provably impossible.  Finally, we have more extensive empirical investigations that illustrate our results.


\subsection{Summary of new and improved results}
First, to demonstrate that the localization we study does not simply occur in all graphs, we present a characterization of localization in PageRank on complete-bipartite graphs (Section~\ref{sec:negative}). Seeded PageRank vectors are combinations of uniform vectors (one for each side of the bipartition) on these graphs. This analysis has two interesting regimes: dense graphs and sparse graphs.
We show that in both of these cases attaining non-trivial accuracy in the 1-norm requires dense PageRank solutions, whereas they localize in the 2-norm.
We conclude that the 1-norm is the most appropriate norm to use in studying localization in PageRank, in the sense that using other norms to measure error can mistakenly report a sparse approximation of a near-uniform vector as being accurate.

Second, we sharpen our result from~\citet{nassar2015strong} in a way that results in a minor improvement to the bound (Theorem~\ref{thm:non-zeros}). The key to this result is a particular type of skewed degree sequence (Definition~\ref{def:degree-sequence}), and we present a new analysis of a number of real-world networks to determine where our theory makes non-trivial predictions (Section~\ref{sec:exp:real-world}). In addition, in this paper we are able to further improve our previous bound under the assumption that a graph is undirected (Theorem~\ref{thm:non-zeros2}).

Third, we expand our empirical evaluations of our methodologies and our theoretical predictions on graphs as follows. We increase the size of the synthetic datasets from our previous studies to graphs of size up to $10^9$ nodes (Section~\ref{sec:exp:synthetic}) and use an additional graph generation mechanism to show that our theoretical bounds on localization are robust across different graph sizes and graph generators (Section~\ref{sec:localization-consistency-bisquik}).
Our comparison with empirical evaluations on real-world networks shows that our theoretical bound is likely to significantly underestimate localization compared with real data.
Our comparisons of localization behavior across different graphs with similar degree sequences show that different graphs with the same degree sequence will have nearly the same localization behavior, unless their clustering coefficients are markedly different (Section~\ref{sec:exp:real-world-localization}).

\section{Related work}
The most closely related work is our previous conference paper where we established a set of similar, but weaker, results~\cite{nassar2015strong}, as well as a previous study of the matrix exponential~\cite{gleich2014sublinear}. These all used the same type of skewed-degree sequence to show localization.

For a substantially weaker notation of approximation, seeded PageRank was also known to be localized as shown in~\citet{andersen2006-local}.   More specifically, they prove that seeded PageRank $\vx$ on any graph can be weakly approximated with a \emph{constant} number of non-zero entries, and where the approximation $\hvv{x}$ satisfies $\eps \cdot \text{degree}(j)\geq  x_j - \hat{x}_j > 0$ for all entries $j$. A slightly relaxed version of this accuracy requirement that is more relevant to the themes of this paper can be stated as follows: $\hvv{x}$ satisfies a degree-normalized infinity norm,
$\|\mD\inv( \vx - \hvv{x} )\|_{\infty} < \eps$, where $\mD$ is the diagonal matrix of node degrees. We explore the impact of degree-weighting on our more global notions of approximations in Section~\ref{sec:negative}. For the problem of computing local estimates of seeded PageRank vectors,~\citet{lofgren2014fast,lofgren2015bidirectional} show a number of scenarios where it is possible to determine if the $i,j$ entry of $(1-\alpha) (\mI - \alpha \mP)^{-1}$ is above a value such as $1/n$.

Another related idea is to locally compute a \emph{piece} of the global or uniform PageRank vector, which corresponds to setting the reset distribution to the uniform distribution over nodes, for the graph with bounded accuracy. Along these lines,~\citet{bar2008local} proved that entries of the PageRank vector in graphs with bounded max in-degree can be computed by a local algorithm that need not explore all vertices of the graph. Without this assumption, the results usually show that $\Theta(n)$ work is required in many plausible computation scenarios for graph access, see~\citet{bressan2014approximating} for a discussion. For instance,  \citet{borgs2014multiscale} show that
the problem of computing PageRank on all nodes with value above a certain threshold requires $\Theta(n)$ work in the general case. Computing only the top-ranked nodes is no easier as shown by \citet{bressan2011local}, where they establish that any deterministic algorithm for computing a constant number of the top \emph{ranked} nodes for general graphs must explore $\Omega(n)$ nodes of the graph~\cite{bressan2011local}. On the other hand, there are positive results on certain classes of randomized algorithms. For example, \cite{bressan2014approximating} present an algorithm for computing, with probability $(1-\eps)$, a $(1\pm\eps)$-approximation of any single PageRank value in \emph{any} graph using only $O( n^{\tfrac{2}{3}} \log(n)^{\tfrac{1}{3}})$ graph access queries.

We note that the concepts of localization and local algorithms are related to the idea of \emph{sketching} algorithms, which are, in broad terms, small-size approximations of large-size objects such that the approximation preserves properties of interest of the larger object. Frequently, sketching algorithms take the form of using linear projections on an object with super-linear size; for example, the connectivity properties of a graph with $O(n^2)$ edges can be approximated by first computing a sparisfied graph with $O(n)$ edges using a sketching procedure, and then studying the connectivity of the sketched graph~\cite{Ahn:2012:GSS:2213556.2213560}.
If a vector is localized, an algorithm could, in principle, compute a sparse approximation in work that scales with the localization instead of the graph size. Sketching, in this scenario, might either reflect sketching the graph data itself to make localized solutions easy to determine, or sketching a single solution vector. Under our degree-distribution assumption with $p \ge 0.5$, our results imply that the there exists a sketch of a seeded PageRank solution vector that has $O(d^2)$ entries (Theorem~\ref{thm:non-zeros}), where $d$ is the max degree. Moreover, our results imply there is a sketch of the PageRank solution matrix $(1-\alpha)(\mI - \alpha \mP)^{-1}$ with $nd^2$ non-zero entries, compared with the trivial $n^2$ bound.

More generally, localization in PageRank is intimately tied to the decay of entries of functions of matrices. Here, we are concerned with the elements of the matrix $(1-\alpha) (\mI -\alpha \mP)^{-1}$, which corresponds to the function $f(x) = (1-\alpha)/(1-\alpha x)$ applied to $\mP$.
Study of the decay of entries of a function of a banded matrix has a long history, with early research focusing on the decay of off-diagonal entries of the inverse of banded matrices~\cite{demko1977inverses,demko1984decay}. More recent efforts generalize these decay and localization results to other functions of matrices and generalizations of the banded property to more graph theoretic scenarios involving shortest path distances and degrees~\cite{benzi1999bounds,Benzi-2007-decay,Benzi-2013-decay}.
Advances in the analysis of localization of functions of sparse and banded matrices hence has implications for localization behavior of PageRank.

The related work all points toward a potentially useful aspect of our present focus: namely that deterministic computation of uniform PageRank entries in a local manner is impossible in general.  While we do not address the case of uniform PageRank, for the case of seeded PageRank, by making a plausible assumption on the degree distribution, we are able to beat the worst-case $\Theta(n)$ bound. It is possible that similar results would help alleviate the worst-case scenario for uniform PageRank as well.

\section{Our class of skewed degree sequences}\label{sec:exp:real-world}
\label{sec:distribution}
We wish to make a few remarks about the class of skewed degree sequences where our results apply. Our definition of a skewed degree sequence is:
\begin{definition}[Rank-skewed degree sequence] \label{def:degree-sequence}
 A graph $G$ has a ($d,\delta,p)$--rank-skewed degree sequence if
 the maximum degree is $d$, minimum degree is $\delta$, and
 the $k$th largest degree, $d(k)$, satisfies $d(k) \leq \max \left\{ d k^{-p},~\delta \right\}$.
\end{definition}

Perhaps the most well-known example of a skewed degree sequence in real-world networks
is the power-law degree distribution
where the probability that a node has degree $k$
is proportional to $k^{-\gamma}$.
 These power-laws can be related to our rank-skewed sequences with $p = 1/(\gamma-1)$ and $d = O(n^p)$~\cite{avrachenkov2012quick}.

 Our theoretical bounds in Theorem~\ref{thm:non-zeros} show that there are asymptotically $\tilde{O}(d^{1/p})$ non-zeros in a seeded PageRank vector, and so this power-law setting renders our bound trivial with $\Theta(n)$ nonzeros.
 Nevertheless, there is evidence that some real-world networks exhibit our type of skewed degrees~\cite{faloutsos1999power} where the bound is asymptotically non-trivial. Below we explore a dozen real-world datasets to see how well our rank-skewed degree sequence setting models graphs in practice and to understand whether $d^{1/p}$ scales sublinearly in those settings. We find that there are multiple real-world networks where our theory yields non-trivial predictions and find that $0.5 \le p \le 1$ is a reasonable regime.

\begin{figure}
\includegraphics[width=0.33\linewidth]{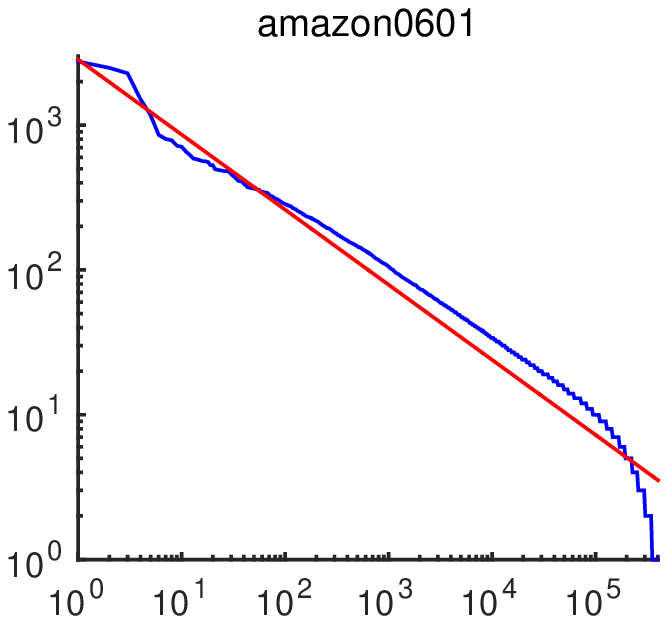}%
\includegraphics[width=0.33\linewidth]{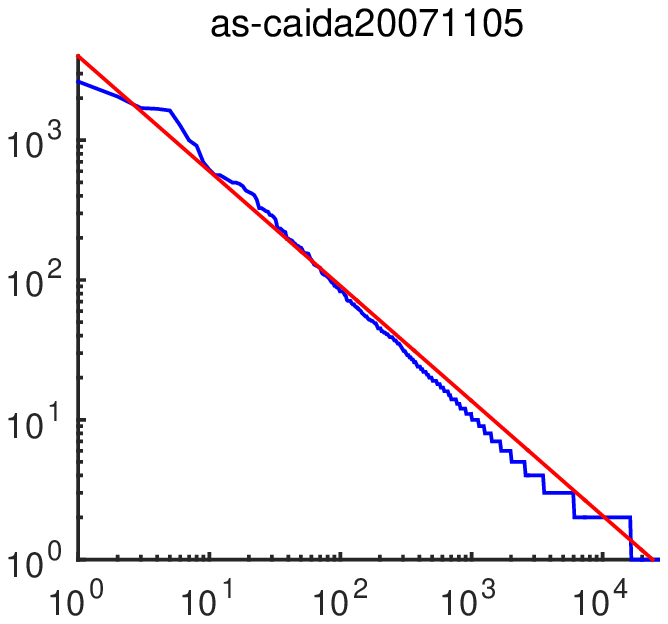}%
\includegraphics[width=0.33\linewidth]{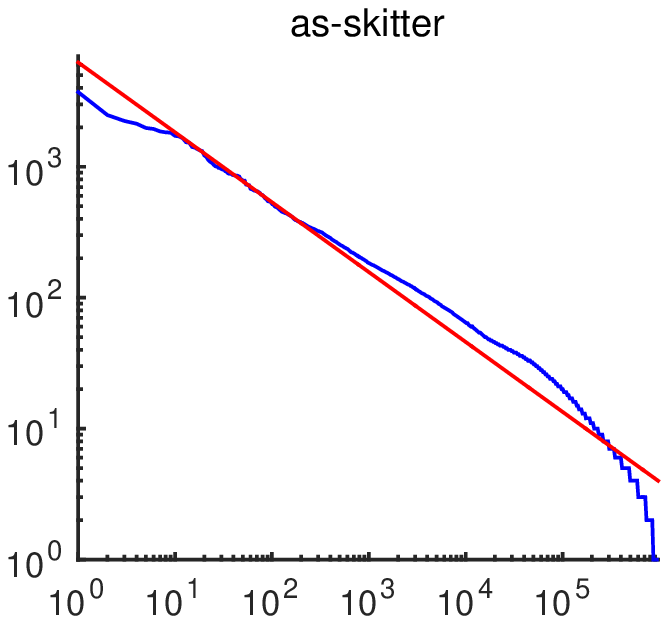}\\
\vspace{5pt}
\includegraphics[width=0.33\linewidth]{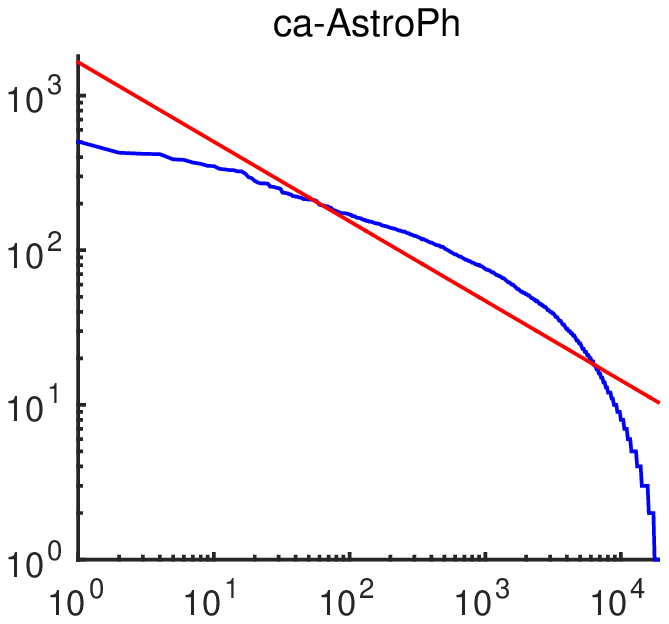}%
\includegraphics[width=0.33\linewidth]{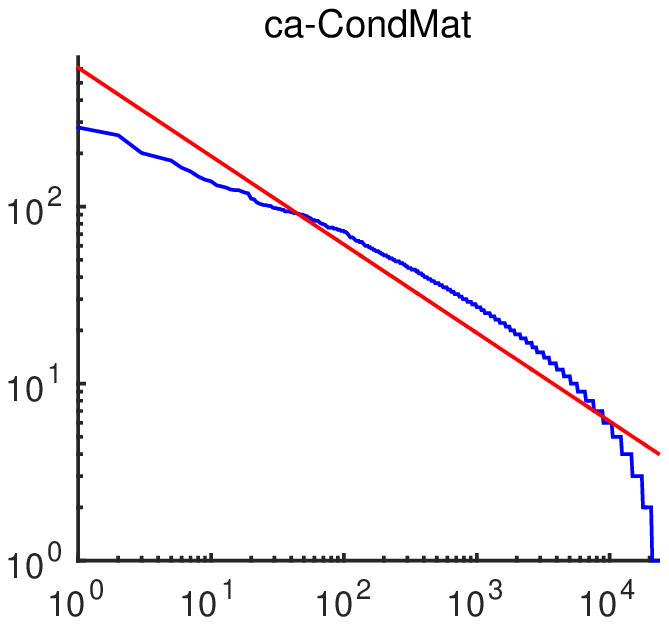}%
\includegraphics[width=0.33\linewidth]{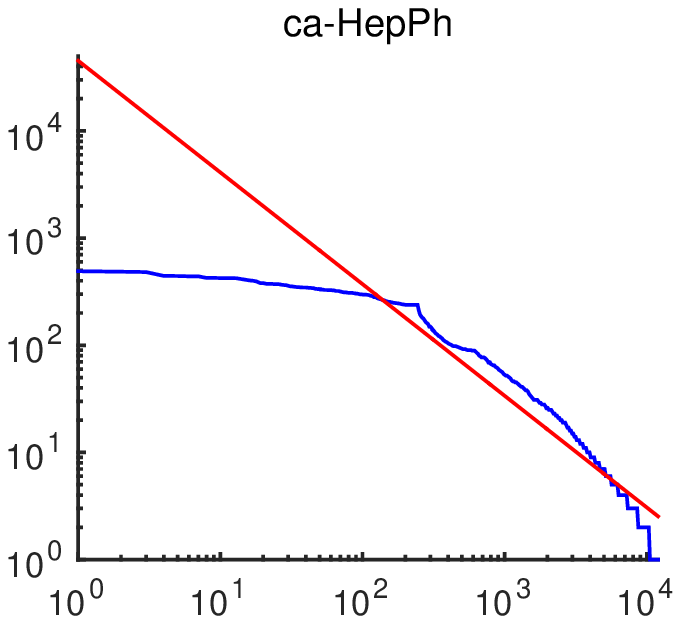}\\
\vspace{5pt}
\includegraphics[width=0.33\linewidth]{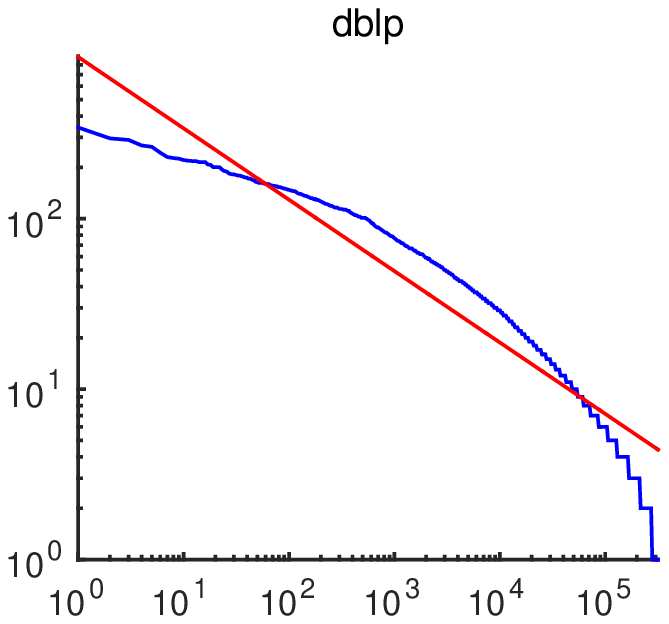}%
\includegraphics[width=0.33\linewidth]{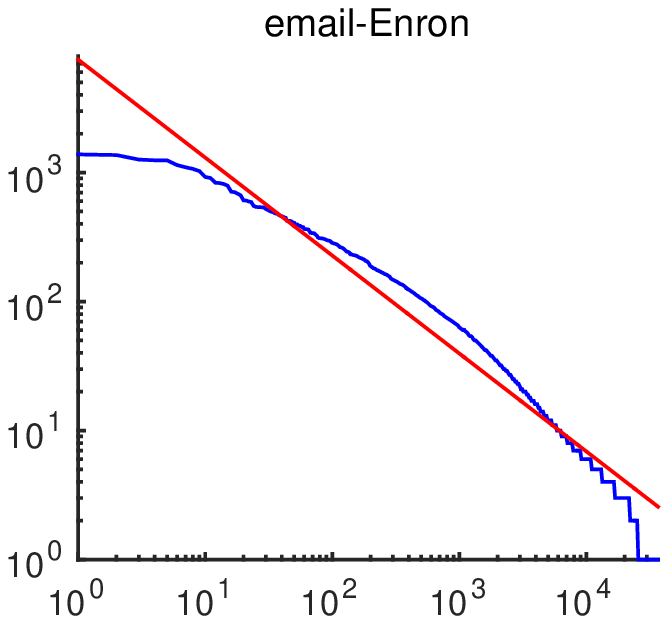}%
\includegraphics[width=0.33\linewidth]{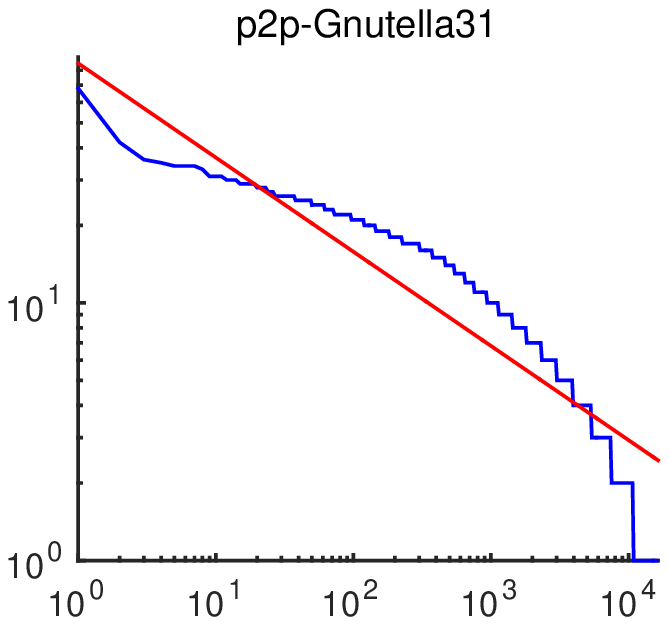}\\
\vspace{5pt}
\includegraphics[width=0.33\linewidth]{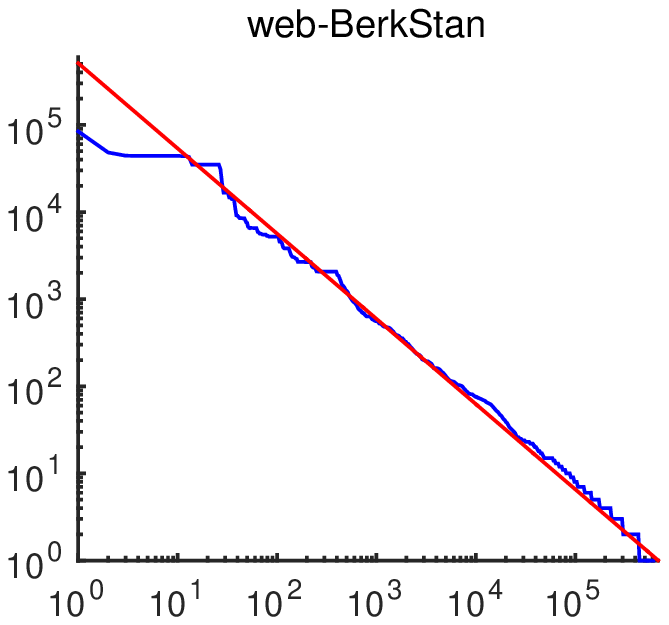}%
\includegraphics[width=0.33\linewidth]{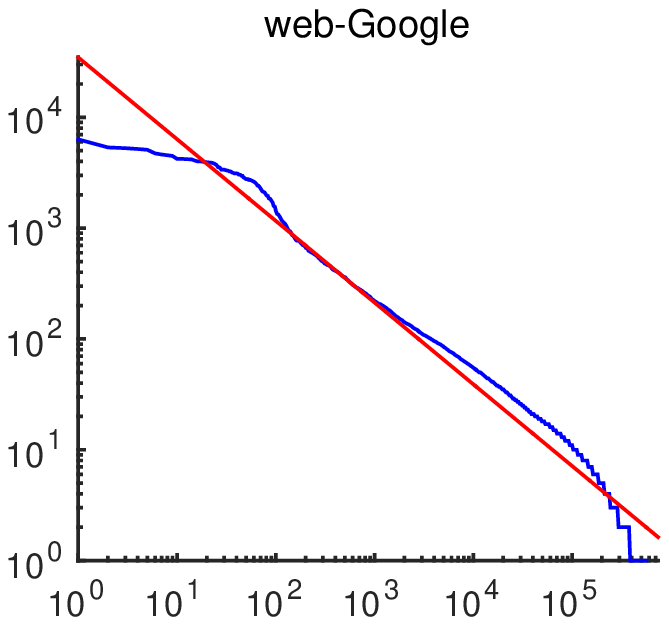}%
\includegraphics[width=0.33\linewidth]{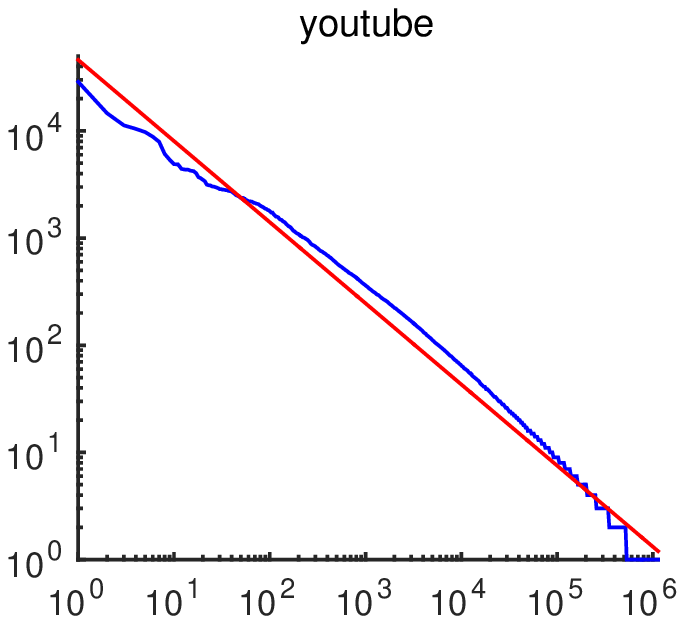}%
\caption{Log-log plots of the degree sequences from real-world networks in terms of degree-rank (horizontal) to degree (vertical). For each dataset, we also plot a best-fit line (red) to show how the actual degree sequence (blue) compares to the rank-skewed degree sequence property that we assume in our theory. Though several datasets have degree sequences that are too lop-sided to match our skewed property (e.g.~the collaboration networks in the middle two rows), a number of other graphs are well-modeled by our skewed degree sequence (the top and bottom rows).
We present the max degree, the slope of each best-fit line, and the size of our theoretical bound for each best-fit line in Table~\ref{tab:real-world-deg-seq}.}\label{fig:realworld}
\end{figure}

\paragraph{Skewed degree sequences in real-world networks.}
To study how practical our skewed degree sequence is, we look at the degree sequences of a variety of real-world datasets. The datasets we study are from the SNAP repository~\cite{snap}. We use collaboration networks from Arxiv Astro Physics, Arxiv Condensed Matter, Arxiv High Energy Physics~\cite{leskovec2005shrinking-diameter}, and DBLP~\cite{yang2015defining}; an email communication network from Enron~\cite{leskovec2009community,klimt2004introducing}; a social network YouTube~\cite{mislove-2007-socialnetworks};  an instance of the peer-to-peer network Gnutella~\cite{ripeanu2002mapping,leskovec2005shrinking-diameter}; autonomous systems graphs Skitter and Caida~\cite{leskovec2005shrinking-diameter}; a co-purchasing network formed from Amazon data~\cite{leskovec2007dynamics}; and web graphs from Berkley-Stanford and Google web domains~\cite{leskovec2009community}. The results of the degree distribution and the best-fit parameters are in Figure~\ref{fig:realworld} and Table~\ref{tab:real-world-deg-seq}.

We observe that the skewed degree sequence property and parameter settings we consider in our theoretical section do describe some of these real-world datasets accurately (for example, the bottom row of graphs in Figure~\ref{fig:realworld}).
However, we also see that many of the real-world graphs are not perfectly modeled by our degree sequence assumption,
and it would require a more nuanced model to accurately describe these degree sequences.
In addition to showing how well our skewed degree sequence describes these real-world networks, we also want to see the extent to which our theoretical bound predicts sublinear localization for the degree sequences of these real-world graphs. With this in mind, Table~\ref{tab:real-world-deg-seq} displays the quantity $\log_n(C_p)$ for each network, where $C_p$ comes from Theorem~\ref{thm:non-zeros}. If this quantity is less than 1, then our theoretical bound gives an asymptotically sublinear prediction for localization.
From Figure~\ref{fig:realworld} and Table~\ref{tab:real-world-deg-seq} we can see that the datasets \texttt{as-caida}, \texttt{web-BerkStan}, \texttt{web-Google}, and \texttt{youtube} all exhibit degree sequences near our skewed degree sequence, and also have a sublinear asymptotic factor $C_p$.


\paragraph{Getting the line of best fit.}
We use RANSAC to perform the line fitting task on the data points~\cite{fischler1981ransac}. In order to get a better fit, we sample 500 data points from the original set that are equally-spaced on a log-log scale. Then we use RANSAC on this subset of the nodes. We do this sampling because the degree sequences we are studying have many more nodes of low degrees than nodes of high degrees, forcing any line fitting strategy to heavily bias the line toward the low degree nodes. By sampling points that are equally spaced on a log-log scale, we instead bias toward a line that fits better on the log scale. We note that similar procedures have been shown to mis-estimate the parameters of a standard power-law degree sequence compared with a maximum likelihood estimator~\cite{clauset2009-powerlaw}. This aspect may merit additional investigation in the future.

\begin{table}[t]
\centering
\caption{
This summary of real-world networks and their degree sequences in the context of our skewed degree sequence model and our theoretical localization bound shows that a number of real-world networks fit the regime of our model where our theory implies sublinear localization.
The column $p$ gives the (negative) slope of the best-fit lines for each dataset (pictured in Figure~\ref{fig:realworld}).
The 3rd column displays the exponent $e$ such that $d=n^e$. The far right column shows how the precise asymptotic factor of our localization bound from Theorem~\ref{thm:non-zeros}, $C_p$, compares to $n$. When $\log_n(C_p)$ is less than 1, this indicates our theoretical localization bound is sublinear in $n$ and hence predicts a PageRank vector that is localized. For example, the network \texttt{web-BertStan} has $\log_n(C_p) < 1$ and is modeled well by our skewed degree sequence (see Figure~\ref{fig:realworld}) and so our theory predicts localization in PageRank in this case.
\label{tab:real-world-deg-seq}}
\begin{tabularx}{0.9\linewidth}{lXXXX}
 \toprule
 data  &  p   &   $\log_n(d)$ &  $ \log_n(C_p) $ & Good fit \textit{\&} localized \\
 \midrule
  \texttt{             ca-HepPh } &  1.04  &  0.66  &  0.63 & No\\
 \texttt{         web-BerkStan } &  0.98  &  0.84  &  0.86 & Yes \\
 \texttt{     as-caida20071105 } &  0.82  &  0.77  &  0.94  & Yes\\
 \texttt{              youtube } &  0.76  &  0.74  &  0.97 & Yes \\
 \texttt{          email-Enron } &  0.76  &  0.69  &  0.91 & No \\
 \texttt{           web-Google } &  0.74  &  0.65  &  0.88  & Yes\\
 \texttt{           as-skitter } &  0.53  &  0.60  &  1.12 & No \\
 \texttt{           amazon0601 } &  0.52  &  0.61  &  1.18 & No \\
 \texttt{           ca-AstroPh } &  0.51  &  0.63  &  1.23 & No \\
 \texttt{           ca-CondMat } &  0.50  &  0.56  &  1.12 & No \\
 \texttt{                 dblp } &  0.47  &  0.98  &  2.09 & No \\
 \texttt{       p2p-Gnutella31 } &  0.37  &  0.43  &  1.19 & No \\
\bottomrule
\end{tabularx}
\end{table}


\section{Negative results for strong localization}\label{sec:negative}

The main results of this paper show that, under our rank-skewed degree distribution assumptions on a network, seeded PageRank vectors can be computed in a local manner with provable accuracy. We now show that there are classes of graphs where seeded PageRank vectors \emph{cannot} be computed in a local manner. This shows that the localization property we study is non-trivial as it is not present in \emph{all} networks. Furthermore, we show that there are natural error measures that will report \emph{good} approximations of a near-constant vector with only a sublinear number of entries. These error measures are therefore not appropriate for detecting localization in the solution vectors.


In the rest of this section, we discuss when seeded PageRank vectors \emph{cannot} be locally approximated. Formally, a vector $\vx$ of length $n$ can be locally approximated if we can find a vector $\hvv{x}$ with $o(n)$ non-zero entries that is an $\eps$-accurate approximation $\vx$.
To find a class of graphs with this property, it suffices to consider
 complete-bipartite graphs ($K_{k,n-k}$) because their structure and symmetry enable explicit computation of seeded PageRank vectors.
 Once we construct these PageRank vectors, we then
study the sparsity of the sparsest $\eps$-accurate approximations when approximation error is measured using the 1-norm, 2-norm, and degree-normalized versions of the 1- and 2-norms.

\begin{table}
\caption{Can seeded PageRank be locally approximated on complete-bipartite graphs $K_{k,n-k}$? The answer depends on how $k$ scales with $n$ and how we evaluate norms. When $k = O(1)$ then the graph is sparse whereas when $k = \Theta(n)$, then the graph is dense. The table summarizes our finding that when error is measured using the 2-norm, all seeded PageRank vectors are can be locally approximated -- even though we prove such vectors are essentially uniform distributions with no concentration in the values (Theorem~\ref{thm:general-delocalized-bipartite}). This shows that only $1$-norms are appropriately sensitive to our type of localization. }
\label{tab:delocalization}
\centering
 \begin{tabularx}{0.8\linewidth}{lXX}
 \toprule
 Graph & Error measure & Local approx. \\
 \midrule
  sparse $K_{k,n-k}, k = O(1)$ & $~\| \vx - \hvv{x} \|_1$ & No \\
  sparse $K_{k,n-k}, k = O(1)$ & $~\| \mD\inv (\vx - \hvv{x}) \|_1$ & No \\
  sparse $K_{k,n-k}, k = O(1)$ & $~\| \vx - \hvv{x} \|_2$ & Yes \\
  sparse $K_{k,n-k}, k = O(1)$ & $~\| \mD\inv (\vx - \hvv{x}) \|_2$ & Yes \\
  dense $K_{k,n-k}, k = \Theta(n)$ & $~\| \vx - \hvv{x} \|_1$ & No \\
  dense $K_{k,n-k}, k = \Theta(n)$ & $~\| \mD\inv (\vx - \hvv{x}) \|_1$ & Yes \\
  dense $K_{k,n-k}, k = \Theta(n)$ & $~\| \vx - \hvv{x} \|_2$ & Yes \\
  dense $K_{k,n-k}, k = \Theta(n)$ & $~\| \mD\inv (\vx - \hvv{x}) \|_2$ & Yes \\
  \bottomrule
 \end{tabularx}

\end{table}

\paragraph{Summary of negative results.}
Our results on localization behavior on complete bipartite graphs all rely on our explicit construction of a seeded PageRank vector.
The details of that construction, and subsequent proofs of our localization bounds, are tedious, and outside the main interests of this paper, so we include them in Appendix~\ref{sec:app}, but we briefly overview the implications here and direct the interested reader to specific sections in the appendix with parantheticals.

For a complete-bipartite graph on $n$ nodes with partitions of arbitrary sizes (denoted $k$ and $n-k$), our analysis (Section~\ref{sec:complete-bipartite}) shows that all seeded PageRank vectors consist of the seed node plus two scaled vectors of all 1s: nodes in the same partition as the seed all have the same PageRank value, and nodes outside the seed's partition all have the same PageRank value (see Theorem~\ref{thm:general-delocalized-bipartite} for a precise statement). In short, complete-bipartite graphs have seeded PageRank vectors that are essentially two uniform distributions pasted together.

We use this construction to show that, for any complete-bipartite graph, all nodes have non-local seeded PageRank vectors when error is measured in the 1-norm (Section~\ref{sec:delocalization-1norm}). More concretely, we demonstrate that, in this setting, seeded PageRank vectors require
$\Theta(n)$ nonzeros to attain a 1-norm accuracy of $\eps$, where  $n$ is the number of nodes in the graph.
We present our formal result here, and provide details in the appendix.
\begin{proposition}\label{thm:localized-1norm}
Let $\mP$ be the random-walk transition matrix of an $n$-node complete-bipartite graph, and let $j$ be the index of any node in the graph.
Fix a value $\alpha \in (0,1)$, and a desired accuracy $\eps < \alpha^2/(1+\alpha)$.
Then the number of nonzeros required to approximate the seeded PageRank vector $(1-\alpha)(\mI-\alpha\mP)\inv\ve_j$ with a 1-norm accuracy of $\eps$ is bounded below by
$(1 - \eps (1+\alpha)/\alpha^2)n$.
\end{proposition}
This result on complete-bipartite graphs generalizes our analysis of star graphs in our previous work~\cite{nassar2015strong}.

In Section~\ref{sec:delocalization-1norm} we also consider localization of seeded PageRank vectors when error is measured in the degree-normalized 1-norm, i.e. $\| \mD\inv( \vx - \hvv{x} )\|_1$. We demonstrate that in $K_{k,n-k}$ graphs that have partitions of the same size, seeded PageRank can be approximated with a sublinear number of nonzeros. Such instances of $K_{k,n-k}$ are dense, having $O(n^2)$ edges.
In contrast, we then show that in \emph{sparse} $K_{k,n-k}$ graphs, i.e. complete-bipartite graphs in which one partition is of constant size, seeded PageRank vectors are de-localized even when error is measured as $\| \mD\inv( \vx - \hvv{x} )\|_1$.

In Section~\ref{sec:local-2norm} we move on to consider localization in these same seeded PageRank vectors when error is measured as $\|  \vx - \hvv{x} \|_2$ and as $\| \mD\inv( \vx - \hvv{x} )\|_2$. In both cases we show that seeded PageRank is always localized for arbitrary complete-bipartite graphs.

We end with a toy example on a star graph to explore differences in the localization behavior of an approximate PageRank vectors and its related residual vector (Section~\ref{sec:local-residuals}). A number of fast algorithms for producing sparse approximations of PageRank rely on the residual of the PageRank linear system to determine when the algorithm has converged to an accurate PageRank approximation. We show that, depending on the norm that is used, it is possible to have a sparse residual with norm less than $\eps$ even when the true solution is de-localized.

\paragraph{Conclusion from negative results.}
We conclude from this study of localization in different norms that the 1-norm is the most effective norm for analyzing localization behavior. This is because the 2-norm and degree normalized variations all identify nearly uniform distributions as being localized. We therefore recommend that future studies of localization in seeded PageRank and algorithmic termination criteria for approximating seeded PageRank use the 1-norm to measure error. Alternatively,
future work could explore the utility of measuring relative error, i.e.~analyzing $\|  \vx - \hvv{x} \|_2/\|\vx\|_2$, for example. 


\section{Localization in Personalized PageRank}
\label{sec:localization}
In this section we show for graphs with a rank-skewed degree sequence, \emph{all} seeded PageRank vectors are localized. This result has broad utility because it implies that \emph{every} seeded PageRank vector can be computed locally for every graph in this class.

The bounds we present in this section build on top of
our recent work with seeded heat kernel vectors on graphs with a skewed degree sequence \cite{gleich2014sublinear}. We first adapted that work to obtain localization bounds on seeded PageRank vectors in our subsequent work~\cite{nassar2015strong}.
Here we use similar methodologies as in our previous works, but introduce minor improvements for a sharper overall bound on localization.
Compared to our PageRank localization result in~\cite{nassar2015strong}
we give a tighter upperbound. We then present a second result that further improves our bound by assuming the graph is undirected.
Though we use techniques from our prior work, we provide self-contained proofs in this paper for the convenience of the reader.

\begin{theorem}\label{thm:non-zeros}
Let $\mP$ be a uniform random walk transition matrix of a graph with $n$ nodes and with a $(d,\delta,p)$--rank-skewed degree sequence.
The Gauss-Southwell coordinate relaxation method applied to the seeded PageRank problem $(\mI - \alpha\mP)\vx = (1-\alpha)\ve_s$ from the initial solution $\vvk{x}{0} = (1-\alpha)(\mI + \alpha\mP)\ve_s$
produces an approximation $\vx_{\eps}$ satisfying
$\|\vx - \vx_{\eps}\|_1 < \eps$
having at most $N$ non-zeros in the solution, where $N$ satisfies
\begin{equation}\label{eqn:theoretical_bound}
N = \min\left\{~n~,~1+d+\tfrac{1}{\delta}C_p \left(\tfrac{\alpha^2}{\eps}\right)^{\tfrac{\delta}{1-\alpha}} \right\},
\end{equation}
and where we define $C_p$ to be
\begin{align*}
C_p &:=  d(1 + \log d)& \text{if $p = 1$} \\
&:= d\left( 1+\tfrac{1}{1-p}\left( d^{\tfrac{1}{p}-1} - 1\right) \right) &    \text{otherwise.}
\end{align*}

\end{theorem}
Note that the upper bound $N = n$ is trivial because a vector cannot have more non-zeros than entries. Thus, $d$, $\delta$, $p$, and $n$ must satisfy certain conditions to ensure that inequality~\eqref{eqn:theoretical_bound} is not trivial.
In particular, for values of $p < 1$, it is necessary that $d = o(n^p)$ for inequality~\eqref{eqn:theoretical_bound} to imply that $N = o(n)$. For $p > 1$, the bound guarantees sublinear growth of $N$ as long as
$d = o(n)$. Additionally, the minimum degree $\delta$ must be bounded by $O( \log \log n )$. Thus we arrive at:
\begin{corollary}
	Let $G$ be a class of graphs with degree sequences obeying the conditions of Theorem~\ref{thm:non-zeros} with constant $\delta$ and $d=o(n^{\min (p,1)})$. Then seeded PageRank vectors are localized.
\end{corollary}
We also note that the theorem implies localized seeded PageRank vectors for any graph with a maximum degree $d = O(\log \log n)$.

\textbf{A sharper bound for undirected graphs.}
If we restrict our attention to undirected graphs, we can slightly sharpen our analysis. In our first bound~\eqref{eqn:theoretical_bound} the minimum degree $\delta$ in the exponent of $(\alpha^2/\eps)$ can have a huge impact on the size of the overall bound. Our tighter analysis here changes this factor in the exponent to $\delta-1$. This change affects the size of only the constant in the bound, $(\alpha^2/\eps)^{ (\delta-1)/(1-\alpha)}$, and so it might not seem like much of an improvement.
However, for small values of $\delta$, this can make a large difference. For example, when $\delta = 2$ the improvement makes the large constant in our new theoretical upperbound the square root of the large constant in our old upperbound: $(\alpha^2/\eps)^{2/(1-\alpha)}$ compared to $(\alpha^2/\eps)^{1/(1-\alpha)}$.
We state the result here, and present the proof  following our proof of Theorem~\ref{thm:non-zeros}.

\begin{theorem}\label{thm:non-zeros2}
Let the setting be the same as in Theorem~\ref{thm:non-zeros}, except we assume the graph is undirected and the minimum degree $\delta \geq 2$.
The Gauss-Southwell coordinate relaxation method applied to the seeded PageRank problem $(\mI - \alpha\mP)\vx = (1-\alpha)\ve_s$ from the initial solution $\vvk{x}{0} = (1-\alpha)(\mI + \alpha\mP)\ve_s$
produces an approximation $\vx_{\eps}$ satisfying
$\|\vx - \vx_{\eps}\|_1 < \eps$
having at most $N$ non-zeros in the solution, where $N$ satisfies
\begin{equation}\label{eqn:theoretical_bound2}
N = \min\left\{~n~,~1+d + \tfrac{1}{\delta-1}C_p \left(\tfrac{\alpha^2}{\eps}\right)^{\tfrac{\delta-1}{1-\alpha}} \right\},
\end{equation}
where $C_p$ is the same as in Theorem~\ref{thm:non-zeros}.
\end{theorem}

Note that both upperbounds have a factor $\alpha^2/\eps$ such that if $\alpha^2/\eps < 1$ then the constant attached to $C_p$ becomes very small and the bound approaches $d+1$. This makes sense, because if $\alpha^2 < \eps$, then the initial solution is already $\eps$-accurate, making $d+1$ a tight bound on the number of non-zeros in the solution.

Having stated our main theoretical results on the localization of PageRank, in the remainder of this section we will discuss our skewed degree sequence and present a proof of our results.

\subsection{Deriving the bound}

Getting back to the proof, our goal is an $\eps$-approximation, $\vx_{\eps}$, to the equation
$\left( \mI - \alpha \mP \right) \vx = (1-\alpha)\ve_s$
for a seed $s$. Given an approximation, $\hvv{x}$, we can express the error in terms of the residual vector $\vr = (1-\alpha)\ve_s - (\mI - \alpha\mP)\hvv{x}$ as follows:
\begin{equation}\label{eqn:residual-error}
\vx - \hvv{x} \quad = \quad \left( \mI - \alpha \mP \right)\inv\vr.
\end{equation}
Using this relationship, we can bound our approximation's 1-norm accuracy, $\|\vx - \hvv{x}\|_1$, with the quantity $\tfrac{1}{1-\alpha} \|\vr\|_1$. This is because the column-stochasticity of $\mP$ implies that $\| (\mI - \alpha \mP)\inv \|_1 = \tfrac{1}{1-\alpha}$.
Guaranteeing a 1-norm error $\| \vx - \hvv{x}\|_1 < \eps$ is then a matter of ensuring that  $\| \vr\|_1 < (1-\alpha)\eps$ holds.
To bound the residual norm, we look more closely at a particular method for producing the approximation.

\paragraph{The Gauss-Southwell iteration.}
The Gauss-Southwell algorithm is a coordinate relaxation method for solving a linear system akin to the Gauss-Seidel linear solver.
When solving a linear system, the Gauss-Southwell method proceeds by updating the entry of the approximate solution that corresponds to the largest magnitude entry of the residual, $\vr$.
Hence, one iteration consists of altering only one entry of the solution, plus accessing one node's outlinks. This sparse update makes Gauss-Southwell convenient for seeded PageRank-style problems, and it has appeared numerous times in recent  literature \cite{berkhin2007-bookmark,Bonchi-2012-fast-katz,jeh2003-personalized,mcsherry2005-uniform}.
Here we describe the Gauss-Southwell update as it is used to solve the seeded PageRank linear system.

The algorithm begins by setting an initial solution and computing the associated residual. The standard choice for the initial solution is  $\vvk{x}{0} = 0$, but we are able to sharpen our main theorem from~\cite{nassar2015strong} by choosing the initial solution $\vvk{x}{0} = (1-\alpha)(\mI + \alpha\mP)\ve_s$. Then the corresponding residual is $\vvk{r}{0} = (1-\alpha)\ve_s - (\mI-\alpha\mP)\vvk{x}{0} = (1-\alpha)\alpha^2\mP^2\ve_s$.

In step $k$, let $j=j(k)$ be the entry of $\vvk{r}{k}$ with the largest magnitude, and let $m_k = |\vvk{r}{k}_j|$. Gauss-Southwell updates the solution $\vvk{x}{k}$ and residual as follows:
\begin{align}
\vvk{x}{k+1} &= \vvk{x}{k} + m_k \ve_j \\
\vvk{r}{k+1} &= \ve_s - (\mI-\alpha\mP) \vvk{x}{k+1},
\end{align}
and the residual update can be expanded to $\vvk{r}{k+1}
= \vvk{r}{k} - m_k \ve_j + m_k\alpha \mP \ve_j $.
Since each update to the solution $\vvk{x}{k}$ alters exactly one entry of the vector, the index $k$ is an upper-bound on the number of non-zeros that we add to the initial solution vector $\vvk{x}{0}$. (Note that the initial solution $\vvk{x}{0}$ has $d+1$ nonzero entries.)

 To continue, we show that the residual and solution vector stay nonnegative throughout this process. This fact has been proved in previous applications of Gauss-Southwell (see \cite{Bonchi-2012-fast-katz}, Section 5.2, among others) but we include a proof for the convenience of the reader. First, note that the initial residual, $\vvk{r}{0} = (1-\alpha)(\mI+\alpha\mP+\alpha^2\mP^2)\ve_j$, is nonnegative because it is a sum of nonnegative vectors (since $\mP$ is nonnegative). Now, by way of induction assume that $\vvk{r}{k}$ is nonnegative. Then the largest magnitude entry is positive, $m_k > 0$. Hence, in the update of the residual $\vvk{r}{k+1}
= \vvk{r}{k} - m_k \ve_j + m_k\alpha \mP \ve_j$, subtracting $m_k \ve_j$ is equivalent to setting the $j$ th entry of the previous residual $\vvk{r}{k}$ to zero (which is nonnegative by assumption), and then adding the scaled nonnegative vector $\mP \ve_j$ keeps the new vector $\vvk{r}{k+1}$ nonnegative. By induction, this proves that the residual is always nonnegative. Finally, the solution must also always be nonnegative because all solution updates consists of adding $m_k$, a positive quantity, to an entry of the solution. Since the initial solution $\vvk{x}{0} = (1-\alpha)(\mI + \alpha\mP)\ve_j$ is nonnegative, this completes a proof of the claim.

Since we have shown that the residual is always nonnegative, we know the 1-norm of the residual can be expressed as
$\| \vvk{r}{k+1} \|_1 = \ve^T\vvk{r}{k+1}$, where $\ve$ is the vector of all ones. 
Expanding the residual in terms of the iterative update presented above, we can write the residual norm as $\ve^T\left(\vvk{r}{k} - m_k \ve_j + m_k\alpha \mP \ve_j  \right)$. Then, denoting $\|\vvk{r}{k}\|_1$ by $r_k$, we have the recurrence
$r_{k+1} = r_k - m_k (1-\alpha).$

Next, observe that, since $m_k$ is the largest magnitude entry in $\vvk{r}{k}$, $m_k$ is larger than the average value of $\vvk{r}{k}$. Let $Z(k)$ denote the number of nonzero entries in $\vvk{r}{k}$; then the average value can be expressed as $r_k/Z(k)$. Hence, we have $m_k \geq r_k/Z(k)$, and so we can bound $r_k - m_k(1-\alpha)$ above by $r_k - r_k(1-\alpha)/Z(k) $. Thus, $r_{k+1} \leq r_k\left(1 - (1-\alpha)/Z(k) \right)$, and we can recurse to find:
\begin{equation}
r_{k+1} \leq r_0 \prod_{t=0}^k \left(1 - \tfrac{1-\alpha}{Z(t)} \right),
\end{equation}
where $r_0 = (1-\alpha)\alpha^2$ because $\vvk{r}{0} = (1-\alpha)\alpha^2\mP^2\ve_s$.
Then, using the fact that $\log(1-x)\leq -x$ for $x < 1$, we note:
\begin{equation} \label{eqn:residualexp}
r_{k}  \leq  (1-\alpha)\alpha^2 \prod_{t=0}^{k-1} \left(1 - \tfrac{1-\alpha}{Z(t)} \right)
 \leq  (1-\alpha)\alpha^2 \exp\left( -(1-\alpha)\sum_{t=0}^{k-1} \tfrac{1}{Z(t)} \right).
\end{equation}
To proceed from here we need some control over the quantity $Z(t)$, and this is where our assumptions on the degree sequence enter the proof.
The proof steps below build toward a proof of both Theorem~\ref{thm:non-zeros} and Theorem~\ref{thm:non-zeros2}, so that they apply to graphs that are not necessarily undirected, as in Theorem~\ref{thm:non-zeros2}. Then we explain how to modify the proof steps to obtain the sharper bound of Theorem~\ref{thm:non-zeros2} that applies specifically to undirected graphs.


\subsection{Using the degree sequence}\label{sec:power}
We show that, for a graph with a particular kind of skewed degree sequence, the number of entries in the residual obeys the following:
\begin{equation}\label{eqn:fillin-bound}
Z(t) \leq C_p + \delta t,
\end{equation}
where the term $C_p$ is defined in the statement of Theorem~\ref{thm:non-zeros}. We follow our analysis in~\cite{nassar2015strong} that improved on our
previous approach in~\cite{gleich2014sublinear}.
The bound in~\eqref{eqn:fillin-bound} is proved below, but first we use this upperbound on $Z(t)$ to control the bound on $r_k$.

 For the next step of our proof, we prove the following inequality
 \[
-  \sum_{t=0}^{k-1} \tfrac{1}{Z(t)} \leq -\tfrac{1}{\delta} \log\left(  (\delta k + C_p )/C_p \right)
 \]
 via using a left-hand rule integral approximation. To do this, first we use \eqref{eqn:fillin-bound} to bound $1/Z(t) \geq 1/(C_p + \delta t)$. Then we sum and use the integral approximation:
 \begin{align}
\sum_{t=0}^{k-1} \tfrac{1}{Z(t)} &\geq \sum_{t=0}^{k-1} \tfrac{1}{C_p + \delta t} \\
&\geq \int_0^k \tfrac{1}{C_p + \delta x}  \textrm{d}x,
 \end{align}
which equals $\log\left( (C_p + \delta k) / C_p \right)/\delta$, as claimed.
 Continuing with the residual bound, we plug the bound on
 $\sum_{t=0}^{k-1} 1/Z(t)$
 into \eqref{eqn:residualexp} as follows. We obtain
\[
r_{k}  \leq
(1-\alpha)\alpha^2 \exp\left( -\tfrac{(1-\alpha)}{\delta}\log\left(  \tfrac{\delta k + C_p }{C_p} \right)
\right),
\]
which simplifies to $r_{k} \leq (1-\alpha)\alpha^2 \left( (\delta k + C_p )/C_p\right)^{(\alpha-1)/(\delta-1)}$.
Finally, to guarantee $r_k < \eps(1-\alpha)$, it suffices to choose $k$ so that
$\left(  (\delta k + C_p )/C_p \right)^{(\alpha-1)/\delta} \leq \eps/\alpha^2$. This holds if and only if
$(\delta k + C_p ) \geq C_p\left(\alpha^2/\eps\right)^{\delta/(\alpha-1)}$ holds. This inequality is guaranteed by
$
k \geq \tfrac{1}{\delta}C_p\left(\alpha^2/\eps\right)^{\delta/(1-\alpha)} - \tfrac{1}{\delta}C_p.
$
Thus, Gauss-Southwell will produce an $\eps$-approximation as long as the number of steps used is at least $k = \tfrac{1}{\delta}C_p\left(\alpha^2/\eps\right)^{\delta/(1-\alpha)} - \tfrac{1}{\delta}C_p $.
Note that if $\alpha^2/\eps > 1$ then this number is negative -- this simply means that $k = 0$ steps of Gauss-Southwell suffice to reach the target accuracy. (This occurs when the initial solution $\vvk{x}{0}$ already satisfies the desired accuracy.)

 Each step of Gauss-Southwell introduces at most one non-zero in the solution vector, and we know that the initial solution vector $\vvk{x}{0} = (1-\alpha)(\mI + \alpha\mP)\ve_s$ has no more than $1+d$ non-zeros.
This implies that if $1+d+k < n$, then there is an approximation $\vx_{\eps}$ with $N = 1+d+k < n$ non-zeros. If $1+d+ k\geq n$, then this analysis produces the trivial bound $N = n$.

\paragraph{Proving the degree sequence bound.}
Finally, we justify the inequality in~\eqref{eqn:fillin-bound} used in the proof of Theorem~\ref{thm:non-zeros}.
Recall that $Z(t)$ is the number of nonzeros in the residual after $t$ steps. This means that $Z(0)$ is the sparsity of the initial residual, $(1-\alpha)\alpha^2 \mP^2 \ve_j$. To upperbound $Z(t)$, we will consider the number of nonzeros in this initial residual, as well as the maximum number of new nonzeros that could be introduced in the residual during the $t$ steps of Gauss-Southwell.
Essentially, we upperbound the number of nonzeros in the residual using a sum of the largest degrees in the graph.

To upperbound the number of nonzeros in the vector $\mP^2 \ve_j$, first note that the vector $\mP\ve_j$ can have no more than $d$ nonzero entries (recall that $d$ is the maximum degree in the graph). Thus, $\mP^2\ve_j = \mP(\mP\ve_j)$, which is a weighted sum of no more than $d$ columns of $\mP$, can have no more nonzeros than $\sum_{m=1}^d d(m)$, i.e.~the sum of the $d$ largest degrees in the graph. In every iteration of Gauss-Southwell, no more than $d_{\text{new}}$ new nonzeros can be introduced in the residual, where $d_{\text{new}}$ is the largest degree of any node that has not yet been operated on.
Thus, we can upperbound the sparsity of the residual in step $t$ as follows: $Z(t) \leq \sum_{m=1}^{d+t} d(m)$.
When we substitute the decay bound $d(m) \le dm^{-p}$ into this expression, $d(m)$ is only a positive integer when $ m \leq (d/\delta)^{1/p}$. Hence, we split the summation $Z(t) \leq \sum_{m=1}^t d(m)$ into two pieces,
\[
Z(t) \quad \leq \quad \sum_{m=1}^t d(m)
\quad\leq \quad \left(\sum_{m=1}^{
\floor{ (d/\delta)^{1/p} } } d m^{-p}\right)
+ \left( \sum_{m= \floor{(d/\delta)^{1/p} } + 1 }^t \delta \right) .
\]
We want to prove that this implies $Z(t) \leq C_p + \delta t$. The second summand is straight-forward to majorize by $\delta t$. To bound the first summand, we use a right-hand integral approximation:
\[ \sum_{m=1}^{
\floor{ (d/\delta)^{1/p} } } d m^{-p}
\leq
d\left(1 + \int_{1}^{ (d/\delta)^{1/p}  } x^{-p} \text{d}x \right).
\]
This integral is straightforward to bound above with the quantity $C_p$ defined in Theorem~\ref{thm:non-zeros}. This completes the proof.

\subsection{The undirected case}
Here we describe how to modify the above proof of Theorem~\ref{thm:non-zeros} so that it implies a slightly stronger result, restricted to undirected graphs. (The stronger result is Theorem~\ref{thm:non-zeros2}).

Recall that  the inequality~\eqref{eqn:fillin-bound}, $Z(t) \leq C_p + \delta t$, gives an upperbound on the number of nonzeros in the residual at step $t$. We obtain the upperbound by summing together the $t$ largest degrees in the graph. The term $C_p$ contains the sum of the degrees that are larger than $\delta$ -- this consists of the $\ceil{(d/\delta)^{1/p}}$ largest-degree nodes.
All nodes that are introduced in the residual after these must then have degree $\delta$.

We alter this proof as follows. Although it is true that $\delta$ is the \emph{degree} of any node introduced in the residual after the $\ceil{(d/\delta)^{1/p}}$ nodes of largest degree,
these nodes of degree $\delta$ do not actually introduce $\delta$ new \emph{nonzero entries} in the residual. This is because of the additional assumption that the graph is undirected: any node that is nonzero in the residual ``received" its residual mass because one of that node's neighbors was operated on in a previous iteration of Gauss-Southwell. Thus, when we operate on a particular nonzero node $v$ in the residual and make the neighbors of $v$ become nonzero in the residual, at least one of the neighbors of node $v$ was already nonzero. Hence, the number of nonzeros in the residual does not increase by $d(v)$, but by at most $d(v) - 1$. Hence, in~\eqref{eqn:fillin-bound} we can replace $\delta$ with $(\delta-1)$.
It is straight-forward to follow the rest of the steps of the proof of Theorem~\ref{thm:non-zeros} and replace $\delta$ with $(\delta-1)$ to obtain the sharper inequality in Theorem~\ref{thm:non-zeros2}.

\textbf{Minimum degree of 1.}
We want to remark on the problem that this modification appears to cause in the case that $\delta = 1$.
The issue arises when we try to divide by $\delta-1$, which would equal 0 if we have $\delta = 1$. Because we introduce the notation $\delta$ as the ``minimum degree" in the graph, this issue would seem to be problematic, since many graphs have 1 as their minimum degree. However, in reality the above proofs work for any constant $\delta$ that upperbounds the smallest degree in the graph. Thus, the simplest way around this apparent difficulty is just to set $\delta = 2$ even when the minimum degree is 1.
We mention this simple workaround for the case that $d_{\textrm{minimum}} = 1$ to demonstrate that this common situation does not break our theory. However, we remark that this workaround makes the resulting theoretical bound more loose.

\section{Experiments}\label{sec:experiments}

We address three goals in our experimental evaluations and summarize our findings.
\begin{enumerate}
 \item \emph{Do graphs have localized solutions when our theory predicts they should?} We find that the results from Theorem~\ref{thm:non-zeros} are conservative, but do match features of empirical localization on random graphs with our rank-skewed degree sequences. The experiments are discussed in Section~\ref{sec:exp:synthetic}.
 \item \emph{Are these findings consistent across different choices of graph generators and the random samples?} In short, yes, the findings are robust to looking at multiple samples from the same graph generator and robust to the choice of random graph generator. Further discussion is in Section~\ref{sec:identical-degrees}.
 \item \emph{How does localization in synthetic graphs compare to real-world graphs?} Our synthetic networks exhibit worse localization than real-world samples, and this can be attributed to the lack of local clustering in the types of synthetic graphs we study. See Section~\ref{sec:exp:real-world-localization} for the details.
\end{enumerate}

Overall, the experiments we present in this section show that, although our theoretical localization bound seems to describe PageRank localization behavior more accurately on asymptotically large graphs, there remains a gap between the existing theoretical explanations for PageRank localization and the behavior we see in practice. Nevertheless, the bound we present in this paper brings the literature closer to an explanation of the graph properties responsible for the localization so often observed in seeded PageRank in practice.

\paragraph{Procedure used for generating graphs.}
Here we describe the routines used to generate our synthetic graphs with specified degree sequences in this paper.
As a general pipeline, we construct a rank-skewed degree sequence with minimum degree $\delta = 2$, maximum degree $d=\ceil{\sqrt{n}}$.
Given a parameter $p$ for a decay exponent, we produce the degree sequence $d(k) = \max\{\floor{d k^{-p}},~ \delta\}$ for $k = 1, \cdots, n$.
We then use the Erd\H{o}s-Gallai conditions and the Havel-Hakimi algorithm to check if the generated degree sequence is graphical. Once we obtain a degree sequence that is graphical, we proceed to the graph generation step. The failure of the graphical sequence check is often due to the degree sequence having degree values that sum to an odd number. In all the experiments we have performed, perturbing one of the nodes with degree $\delta$ and increasing it by $1$ was sufficient. We use the Bayati-Kim-Saberi procedure~\cite{bayati2010graphgen} as implemented in the bisquik software package~\cite{bisquik} to generate graphs that have our skewed degree sequence. We also use the Chung-Lu model~\cite{chung2002connected} to produce graphs with the correct degree sequence in expectation.

\paragraph{Computing localization.}
We use the following procedure to compute an $\eps$-accurate seeded PageRank vector to get the minimal number of non-zeros. First, we compute a highly accurate estimate of the true PageRank vector $\vx$. Then we sort the values from largest to smallest and continue adding entries in sorted order to our approximation until the difference between the approximation and the true vector is smaller than $\eps$. (To compute the "true" vector we perform enough iterations of a standard power method for computing PageRank to guarantee $10^{-12}$ accuracy.)
This always produces the smallest possible of non-zero entries that gives an $\eps$-accurate approximation.

\paragraph{Changes from conference version.}
We note that, although this analysis follows a similar experimental setup as in our previous study~\cite{nassar2015strong}, the results presented here are from entirely new sets of experiments, expand on the parameter regimes of our previous work, and include an additional graph generation mechanism to study robustness of the results across different graph generators.

\subsection{Our localization theory in synthetic graphs}\label{sec:exp:synthetic}
\label{sec:exp:theoreticalbound}
\begin{figure}[p]
\begin{tabular}{@{}l@{}c@{}c@{}c@{}}
& $p = 0.5$ & $p=0.75$ & $p=0.95$ \\
\rotatebox{90}{\hspace*{35pt}$\alpha = 0.25$}&
\vspace{0pt}\includegraphics[width=0.32\linewidth]{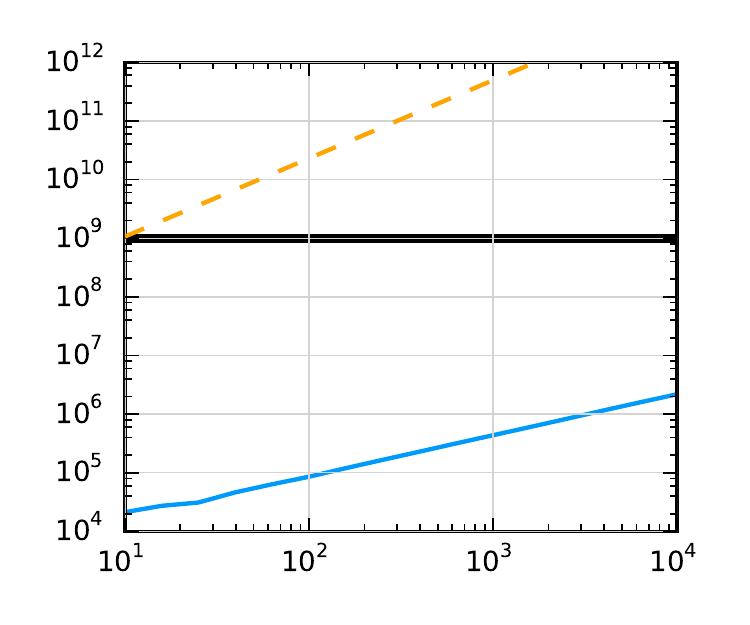} &
\vspace{0pt}\includegraphics[width=0.32\linewidth]{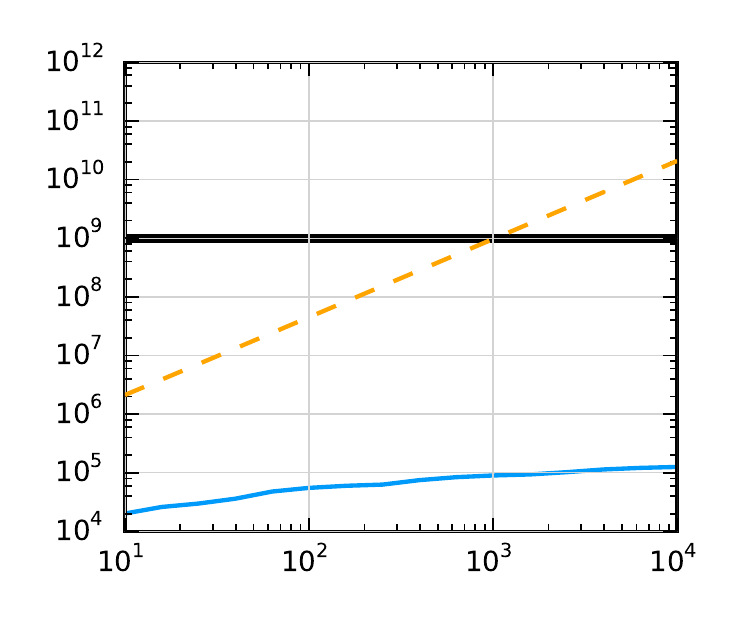} &
\vspace{0pt}\includegraphics[width=0.32\linewidth]{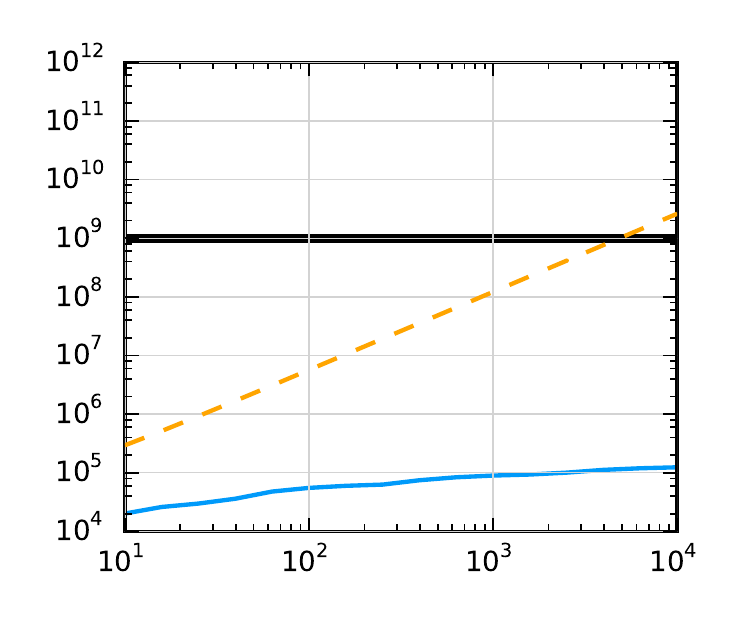}
\\
\rotatebox{90}{\hspace*{35pt}$\alpha = 0.50$}&
\vspace{0pt}\includegraphics[width=0.32\linewidth]{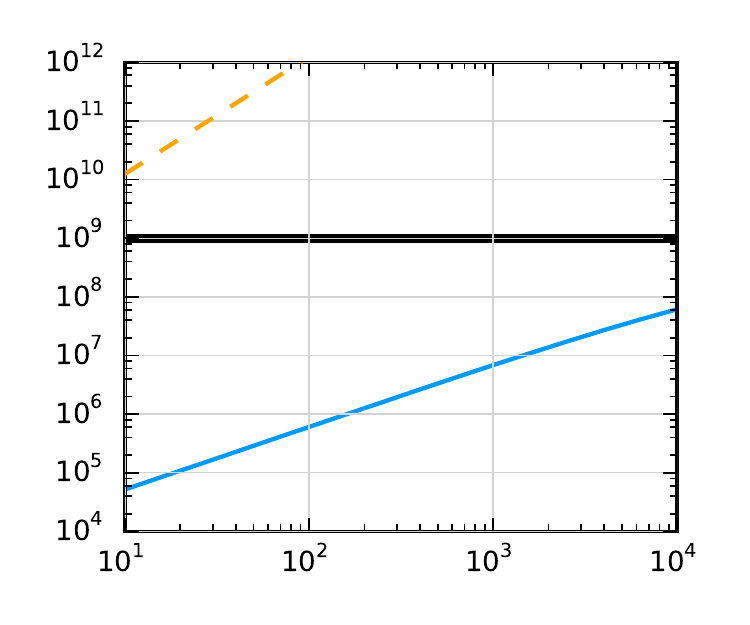} &
\vspace{0pt}\includegraphics[width=0.32\linewidth]{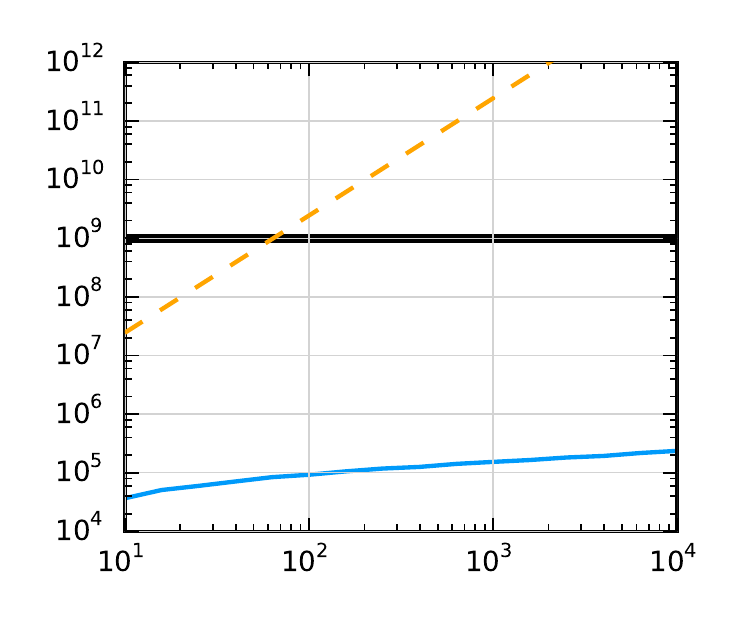} &
\vspace{0pt}\includegraphics[width=0.32\linewidth]{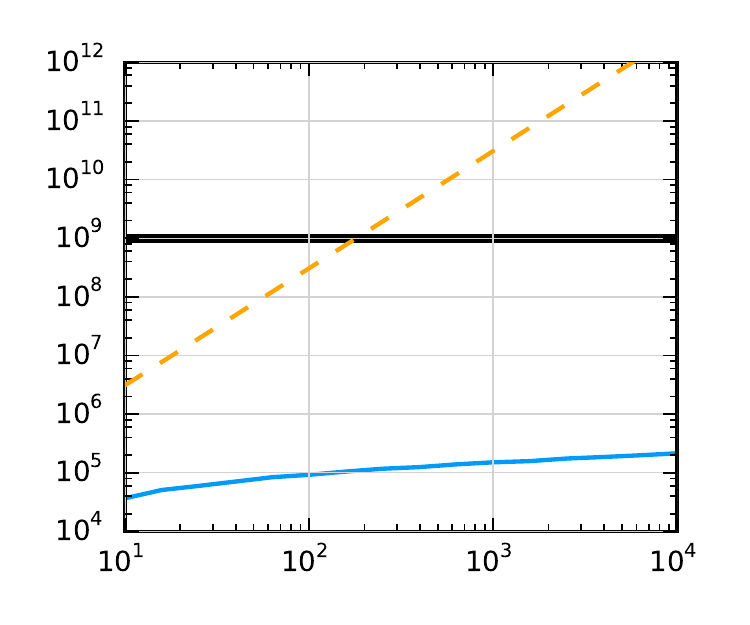}
\\
\rotatebox{90}{\hspace*{35pt}$\alpha = 0.85$}&
\vspace{0pt}\includegraphics[width=0.32\linewidth]{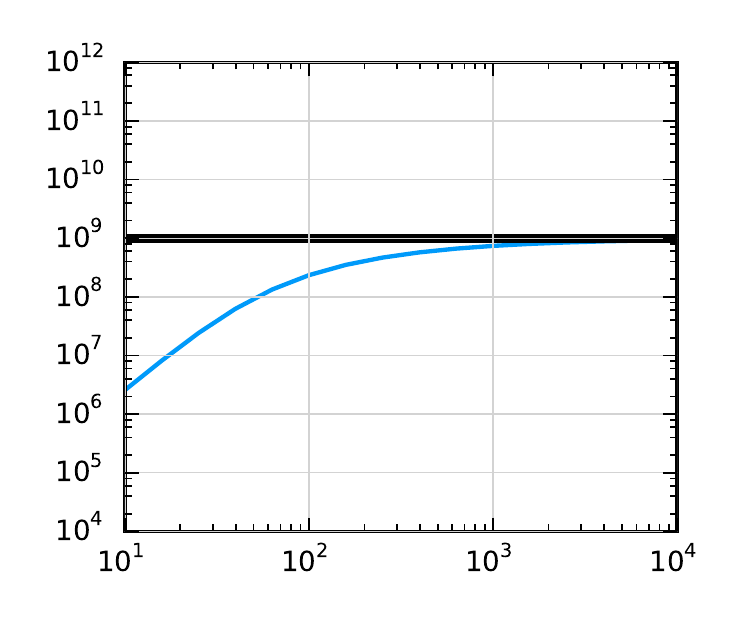} &
\vspace{0pt}\includegraphics[width=0.32\linewidth]{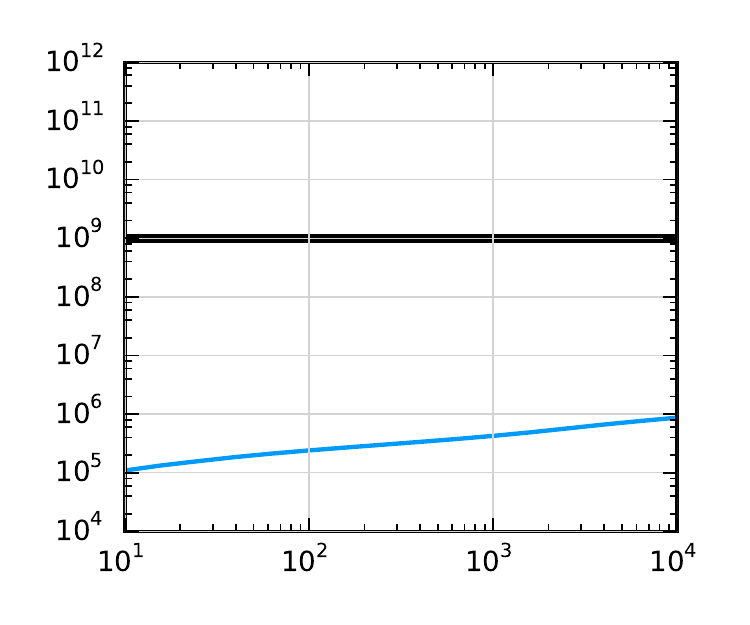} &
\vspace{0pt}\includegraphics[width=0.32\linewidth]{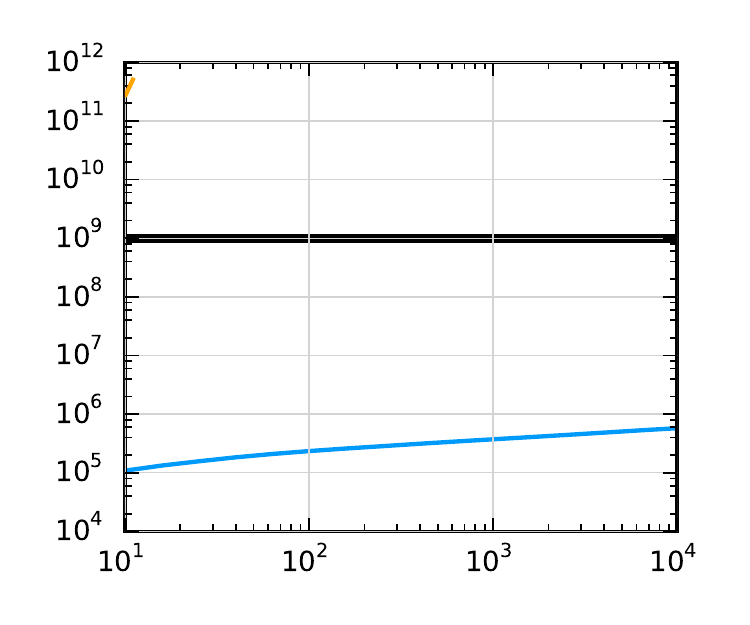}
\\
\end{tabular}
\caption{In each sub-plot the x-axis represents $1/\epsilon$ and the y-axis represents the number of nonzeros present in the PageRank approximation.
The blue curve represents the number of nonzeros in the sparsest PageRank approximation with 1-norm accuracy $\epsilon$. All plots in this figure display results computed on graphs generated with bisquik~\cite{bisquik} with $n = 10^9$, max degree $d = \ceil{n^{1/2}}$, and min degree $\delta = 2$.
The dashed orange line shows the theoretical prediction from Theorem~\ref{thm:non-zeros2}. The thick black line indicates when a solution vector is de-localized, i.e.~has $n$ nonzeros.
We observe far more empirical localization in these graphs than we would predict based on our theory, illustrating that our theory is too coarse. But the we see an increase in localization as $p$ increases, as expected via our theory. In this figure, the size of the graph in all these plots is $10^9$, but the only experiment that delocalizes is $p=0.5$ with $\alpha=0.85$, which uses parameters where our theory does not predict a sublinear bound on the number of nonzeros for any $n$.}\label{fig:theoreticaln1e9}
\end{figure}

In this section we empirically evaluate our theoretical bound on PageRank localization by plotting our bound alongside actual PageRank vectors computed on large synthetic graphs.
For each parameter setting $p = \{ 0.5, 0.75, 0.95\}$ we generate a single graph with $n = 10^9$ nodes and the other parameters as described above. In all experiments we make sure that the largest connected component of the graph is of size at least $0.95 \cdot n$.
Then we compute seeded PageRank on each such graph, seeded on the node of maximum degree, for each parameter setting $\alpha = \{0.25, 0.5, 0.85\}$.
Figure~\ref{fig:theoreticaln1e9} compares our theoretical bound to the actual PageRank localization curve on each different graph. The experiments show that, although the theoretical bound is far from tight, it gives a meaningful prediction of localization for smaller values of $\alpha$ on the graphs other than the densest graph, i.e. the graph generated with $p=0.5$.

We encounter empirical localization in Figure~\ref{fig:theoreticaln1e9} under parameter setting $\alpha = 0.85$: PageRank stays localized for $p = 0.75, 0.95$ where we would not expect localization given our theory.  We believe this occurs for two reasons. First, our theory is obviously too coarse to precisely describe localization. Second, this behavior is a consequence of the way our graphs are constructed. For the parameters in question, the graphs have many long chains of nodes of degree 2 linked to higher degree nodes such that the PageRank diffusion takes many steps to reach a vast portion of the graph. Thus, the diffusion stays localized in these graphs beyond what we would expect.

\subsection{Consistency of localization across samples and graph generators}\label{sec:localization-consistency}\label{sec:localization-consistency-bisquik}\label{sec:identical-degrees}

\begin{figure}
\includegraphics[width=0.33\linewidth]{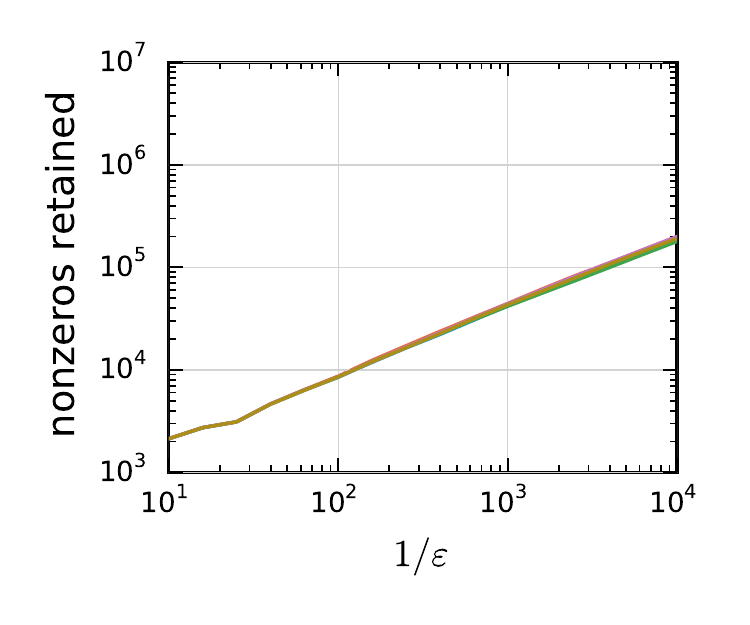}%
\includegraphics[width=0.33\linewidth]{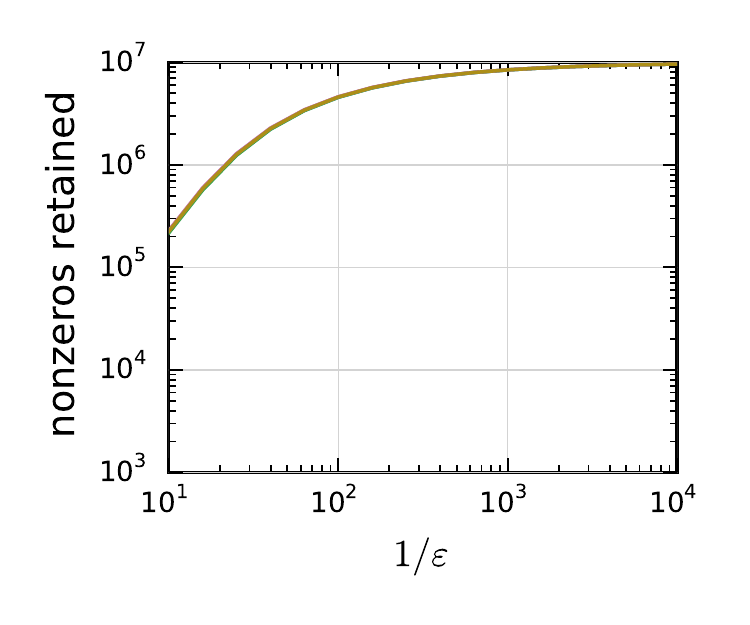}%
\includegraphics[width=0.33\linewidth]{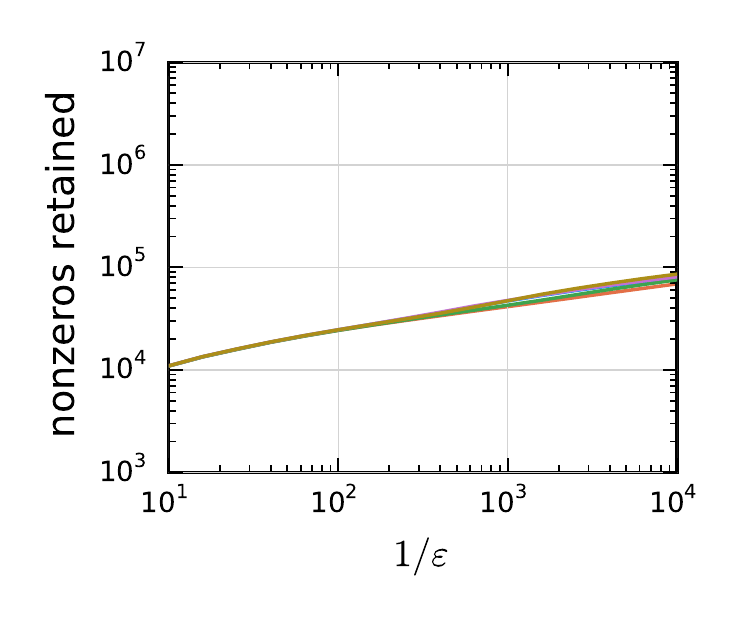}
\caption{These plots display the localization behavior of seeded PageRank vectors computed on multiple different graphs generated by bisquik using the same parameter settings (in particular, the same rank-skewed degree sequence). All graphs used for these plots were generated with $n=10^7$ nodes, maximum degree $d = \ceil{n^{1/2}}$, and minimum degree $\delta = 2$. For each plot, five graphs were generated, and for each graph we computed PageRank seeded on the node of maximum degree.
(\textit{At left}.) PageRank localization curves computed with $\alpha = 0.25$ on five different graphs generated with $p=0.5$. (\textit{Middle}.) PageRank localization curves computed with $\alpha = 0.85$ on the five different graphs with $p=0.5$ used in the left plot. (\textit{At right}.) PageRank localization curves computed with $\alpha = 0.85$ on five different graphs generated with $p=0.95$. All three plots show that PageRank behaves nearly identically on different graphs with the same degree sequence.}\label{fig:localization-consistency}
\end{figure}

\begin{figure}
\includegraphics[width=0.33\linewidth]{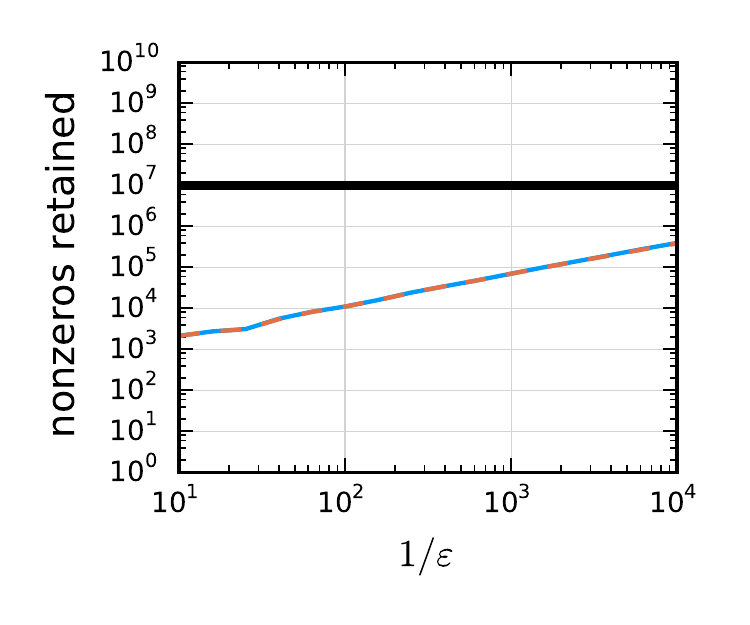}%
\includegraphics[width=0.33\linewidth]{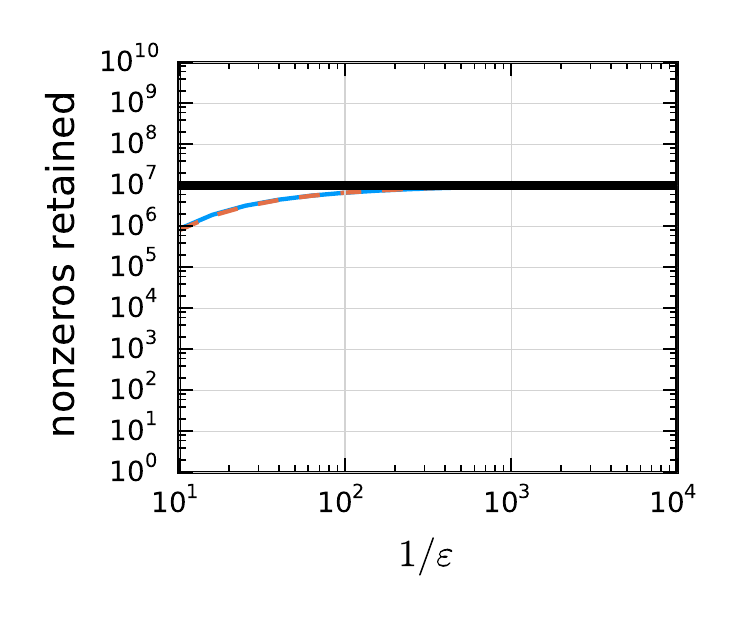}%
\includegraphics[width=0.33\linewidth]{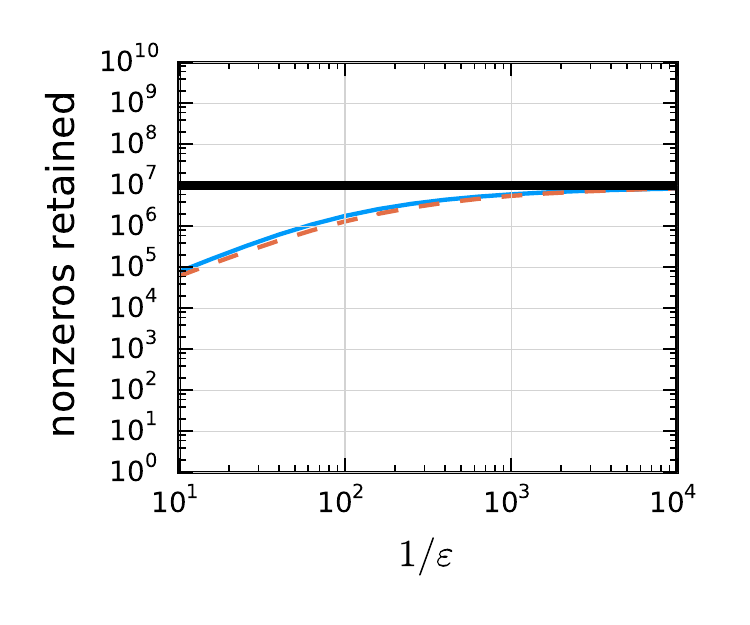}
\caption{
Comparison of PageRank localization on graphs produced by the bisquik procedure (blue) and the Chung-Lu procedure (red-dashed). For this experiment we generated a Chung-Lu graph on $n=10^7$ nodes, set to have our rank-skewed degree sequence in expectation. We then used bisquik to produce a graph with degree sequence identical to that of the Chung-Lu graph. We did this for rank-skewed degree sequences with $p = 0.5$ (the left and center plots) and $p = 0.95$ (the right plot).
These plots show localization curves for PageRank seeded on the maximum degree node of the synthetic graphs. The left plot shows PageRank with $\alpha = 0.25$ while the center and right plots show PageRank with $\alpha = 0.85$. In all cases the localization behavior of PageRank is nearly identical on graphs generated under the Chung-Lu model and the bisquik procedure.
}\label{fig:chung-lu-v-bisquik}
\end{figure}

We would like to know whether the seeded PageRank localization behavior changes significantly on different graph outputs, even though the graphs have the same, or roughly the same, degree sequence.
To explore this question, we use bisquik~\cite{bisquik} to generate multiple graphs with identical parameter settings and observe PageRank localization on each such instance. We also compare bisquik graphs with an exact degree sequence to Chung-Lu random graphs where the degree sequence occurs in expectation.
Figure~\ref{fig:localization-consistency} shows plots of localization curves for five random samples of a bisquik graph with the same degree sequence. Figure~\ref{fig:chung-lu-v-bisquik} compares bisquik to Chung-Lu. There is only minimal variation in the behavior of the PageRank localization across the different graphs and between Chung-Lu and bisquik. These results show that the empirical behavior can be inferred from just a single sample and there do not appear to be extreme sensitivities or variations between samples or graph generation schemes.

\subsection{Localization in real-world and synthetic graphs}
\label{sec:exp:real-world-localization}

\begin{figure}
\begin{minipage}{0.33\linewidth}
\includegraphics[width=\linewidth]{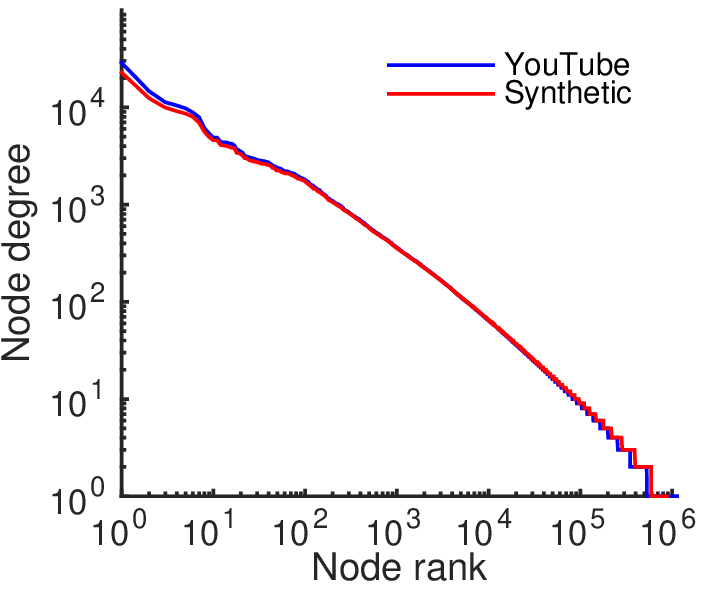}
\end{minipage}%
\begin{minipage}{0.33\linewidth}
\includegraphics[width=\linewidth]{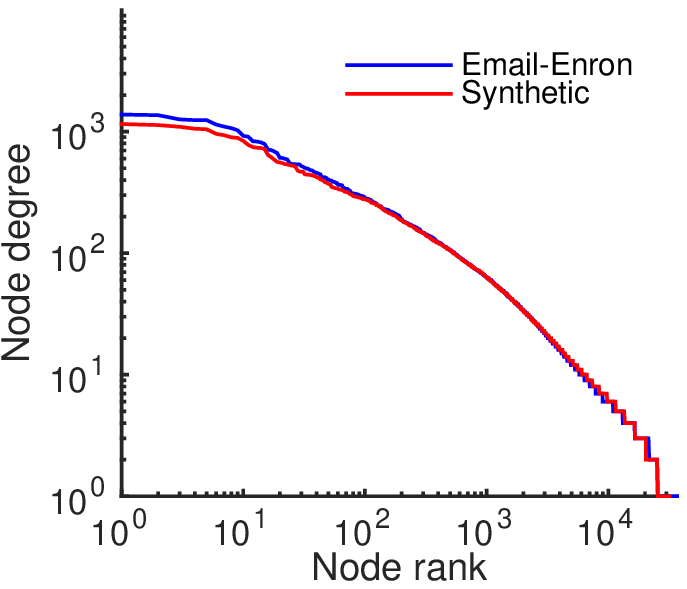}
\end{minipage}%
\begin{minipage}{0.33\linewidth}
\includegraphics[width=\linewidth]{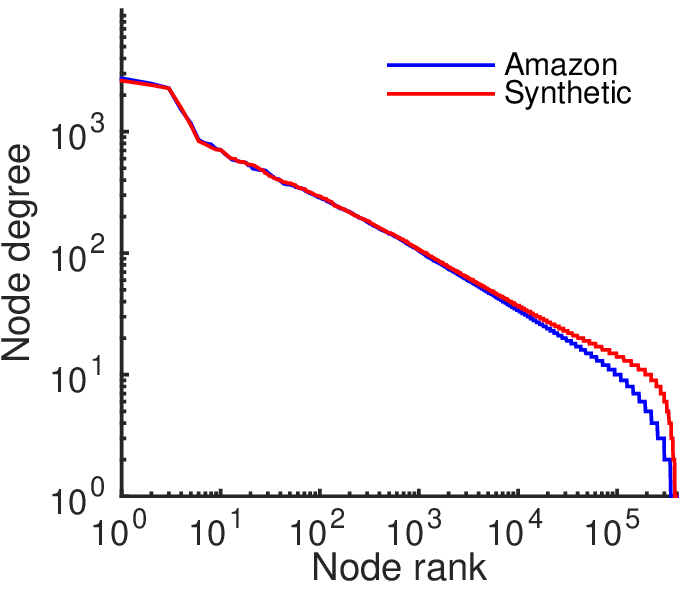}
\end{minipage}%
\\
\begin{minipage}{0.33\linewidth}
\includegraphics[width=\linewidth]{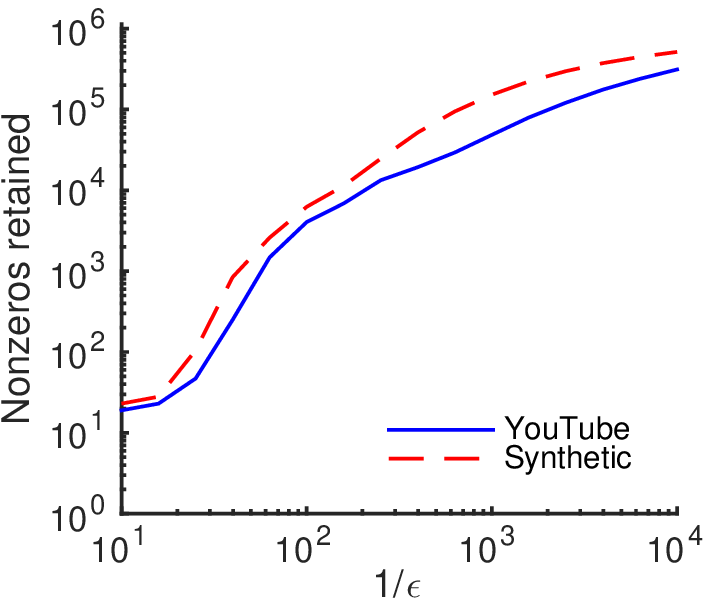}
\end{minipage}%
\begin{minipage}{0.33\linewidth}
\includegraphics[width=\linewidth]{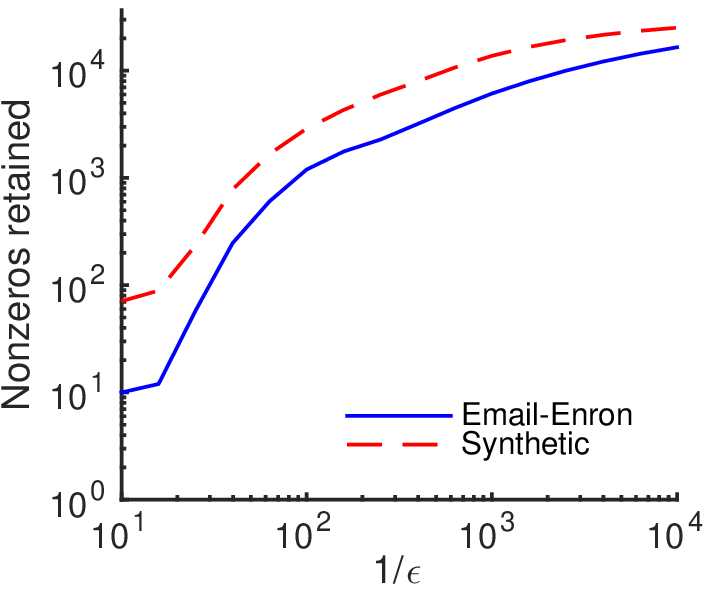}
\end{minipage}%
\begin{minipage}{0.33\linewidth}
\includegraphics[width=\linewidth]{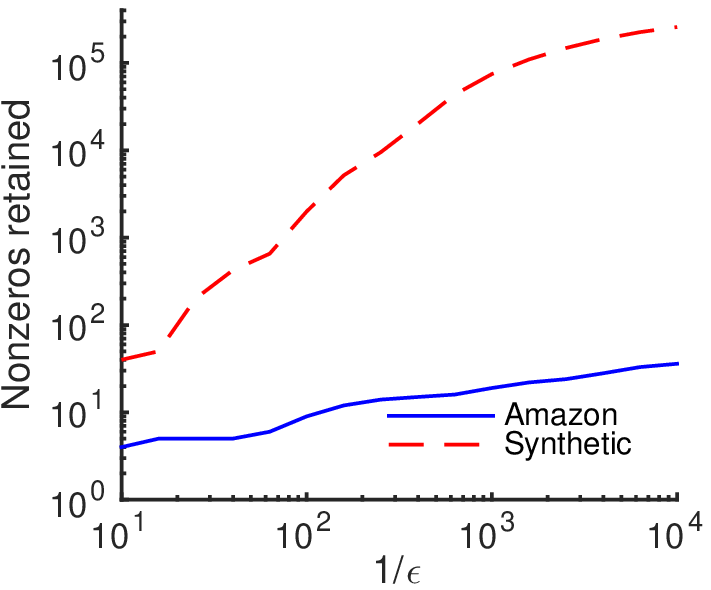}
\end{minipage}%
\\
\begin{minipage}{0.33\linewidth}
\includegraphics[width=\linewidth]{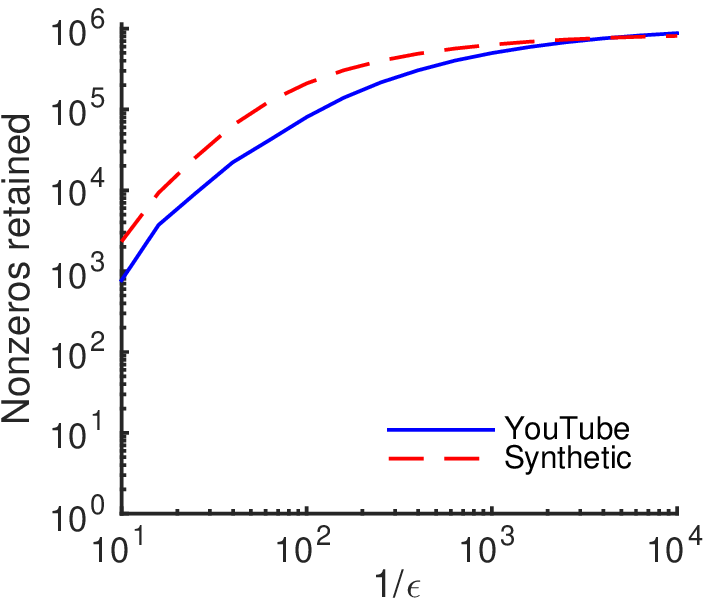}
\end{minipage}%
\begin{minipage}{0.33\linewidth}
\includegraphics[width=\linewidth]{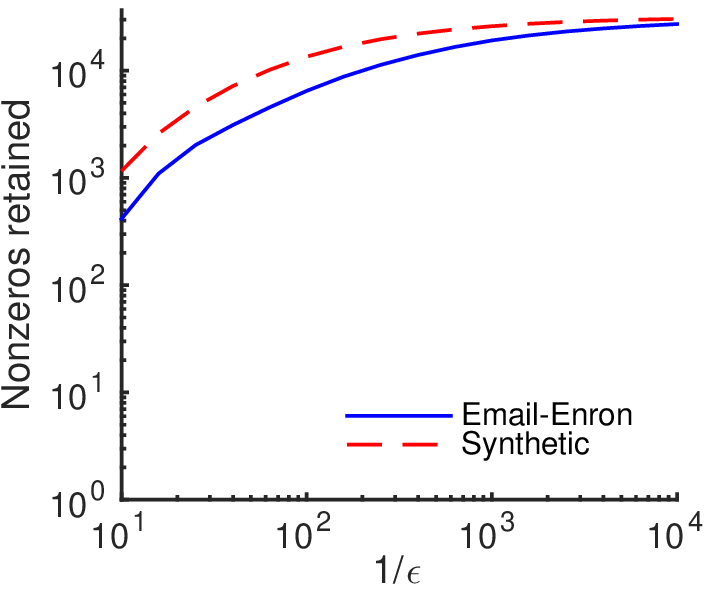}
\end{minipage}%
\begin{minipage}{0.33\linewidth}
\includegraphics[width=\linewidth]{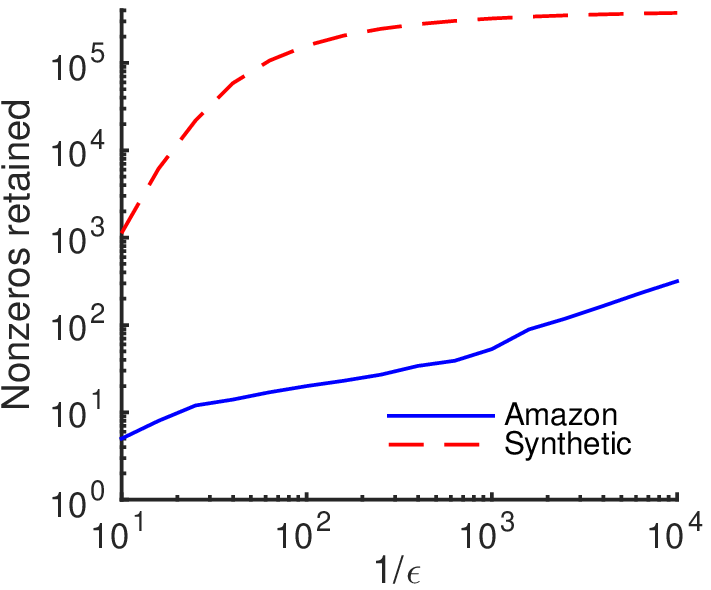}
\end{minipage}%

\begin{minipage}{0.33\linewidth}
\includegraphics[width=\linewidth]{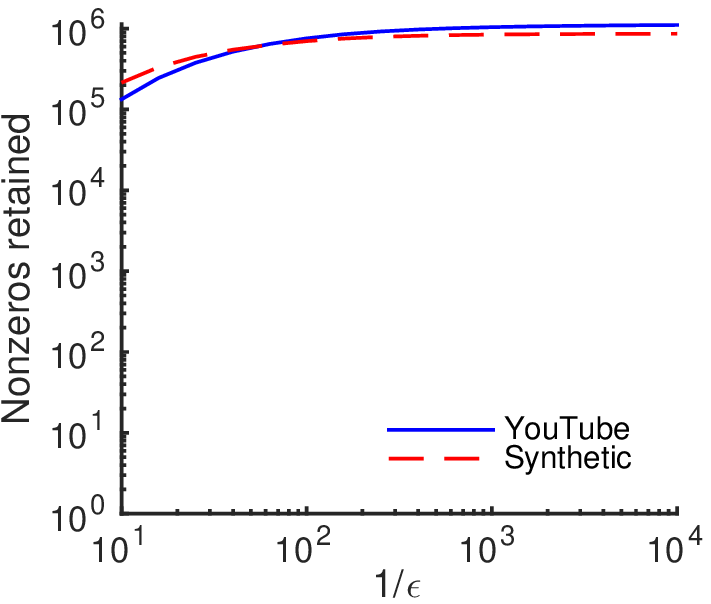}\subcaption{YouTube}\label{fig:a}
\end{minipage}%
\begin{minipage}{0.33\linewidth}
\includegraphics[width=\linewidth]{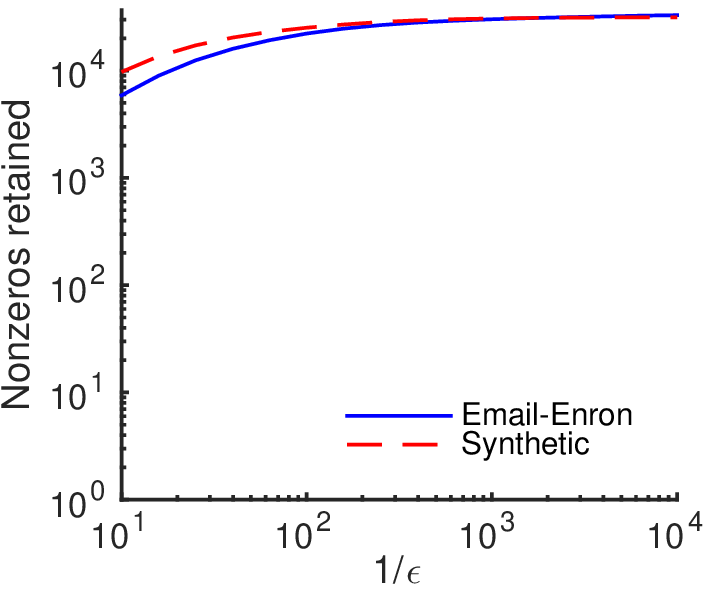}\subcaption{Email-Enron}\label{fig:b}
\end{minipage}%
\begin{minipage}{0.33\linewidth}
\includegraphics[width=\linewidth]{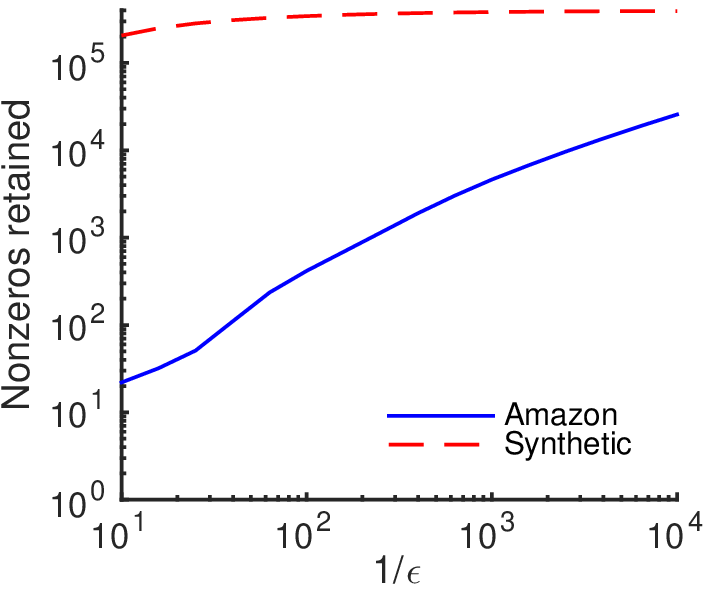}\subcaption{Amazon}\label{fig:c}
\end{minipage}%
\\


\caption{Comparison of PageRank localization in YouTube (\ref{fig:a}), Enron-Email (\ref{fig:b}), and Amazon0601 (\ref{fig:c}) datasets and synthetic graphs generated to have similar degree sequences via Chung-Lu. (\textit{Top Row}) The degree sequences of the three networks and their synthetic counterparts are nearly identical. The other plots show localization in seeded PageRank vectors, seeded on the nodes of maximum degree in the three networks and their synthetic counterparts, for different values of $\alpha$. We use the values $\alpha = 0.25$ (\textit{Second Row}), $\alpha = 0.5$ (\textit{Third Row}), and $\alpha = 0.85$ (\textit{Last Row}). Columns 1 and 2 (YouTube and Enron) show similar localization between the synthetic and real-world graphs. The third column (Amazon) provides an example of two graphs that have nearly identical degree sequences yet starkly different localization. Namely, the real-world graph has substantially more localization than the synthetic graph. We study clustering coefficients in the text to explain this difference.} \label{fig:amazon}\label{fig:enron}\label{fig:youtube}
\end{figure}

In the next set of experiments, we continue to explore how much PageRank localization behavior can change across different graphs that have similar degree sequence. For these experiments we use three real-world datasets (YouTube, Email-Enron, and Amazon, from Section~\ref{sec:exp:real-world}) and, for each of the datasets, we use the Chung-Lu model to generate a synthetic graph with a degree sequence that is nearly identical to that of the real-world network.  The output of the Chung-Lu procedure is not guaranteed to have exactly the degree sequence specified. With this in mind, in Figures~\ref{fig:a},~\ref{fig:b},~\ref{fig:c} (top row) we display the degree sequence of each of the real-world networks alongside the degree sequence of the Chung-Lu graph we produced to mimic the original degree sequences.   We then compute PageRank seeded on the maximum degree node for each graph, with parameter values $\alpha = \{0.25, 0.5, 0.85 \}$. The results are shown in the remaining three rows of those figures.

These experiments show two graphs, YouTube and Email-Enron, where the localization in the real-world and synthetic graph behave in closely related ways. However, for the Amazon graph, the real-world network exhibits starkly more localization. While a precise explanation of the difference is beyond the scope of this paper, we believe this is due to the lack of clustering structure in synthetic networks such as Chung-Lu~\cite{Kolda-2014-BTER} (where each edge is generated independently), and the presence of local clustering structure in real-world data (where there are far more locally dense structures such as triangles than would be expected). Local clustering structure can be quantified by looking at the clustering coefficient of the network, which is the number of triangles (3-cliques) divided by the number of wedges (paths of length 2). Local clustering structure can only \emph{improve} localization because it concentrates the PageRank mass in a smaller region of the graph. In fact, large clustering coefficients and highly-skewed degree sequences can be formally shown to produce vertex neighborhoods with small conductance in a graph~\cite{Gleich-2012-neighborhoods,Yin-2017-local-motif}.

Table~\ref{tab:chung-lu-clones} shows that the graphs with localization behavior closely related to that of their synthetic counterparts (i.e. YouTube and Email-Enron) have clustering coefficients that are also not too different from that of their counterparts. YouTube's global clustering coefficient is nearly the same as its Chung-Lu partner's, Enron's coefficient is a factor of 3 larger than its counterpart's, but Amazon's clustering coefficient is almost 10,000 times larger than its Chung-Lu clone's coefficient. The high level of clustering in the Amazon graph is responsible for the high level of localization we observe in its seeded PageRank vectors (apparent in Figure~\ref{fig:c}); in contrast, the Amazon Chung-Lu clone has almost no clustering, and thus we witness its seeded PageRank vector delocalize rapidly.

\begin{table}[h!]
\centering
\caption{
Each real-world dataset has a synthetic counterpart created via Chung-Lu. Despite having nearly identical degree sequences, the synthetic counterparts can have clustering behavior that differs, sometimes markedly, from the real-world graph. This table shows that the synthetic graphs have lower clustering coefficients than their real-world counterparts, with a particularly sharp difference for the Amazon graph. This difference in clustering behavior likely explains the difference in localization behavior observed in Amazon (Figure~\ref{fig:c}), compared to that of YouTube (Figure~\ref{fig:a}) and Email-Enron (Figure~\ref{fig:b}).
\label{tab:chung-lu-clones}}
\begin{tabularx}{0.8\linewidth}{lXX}
 \toprule
 data  &  clustering coefficient &  clustering coefficient of Chung-Lu sample \\
 \midrule
 \texttt{              youtube } &  0.019  &  0.015  \\
 \texttt{          email-Enron } &  0.26  &  0.085   \\
  \texttt{           amazon0601 } &  0.50  &  0.00054  \\
\bottomrule
\end{tabularx}
\end{table}

\section{Discussion and Future Work}
\label{sec:conclusions}

We have studied localization behavior in approximate PageRank vectors and its relationship both to the degree sequence of the underlying graph and the norm used to measure the error of the approximation.
In particular we proved that,  when error is measured in the 1-norm, seeded PageRank is always de-localized for complete-bipartite graphs, whereas for graphs with a rank-skewed degree sequence it is always localized. Our experiments show that, although this skewed degree sequence does accurately model some real-world networks, there are also many real-world datasets with degree sequences that do not decay sufficiently rapidly for our theory to apply. Thus, there remains an intermediate regime of degree sequences which neither our de-localization theory nor our localization theory describes.
To extend our theory into this regime, we plan to extend our negative results on complete-bipartite graphs to slight edge-perturbations of the highly structured complete-bipartite graphs. This would make progress toward closing the gap between complete-bipartite graphs and graphs with our skewed degree sequence.

Additionally, we are investigating extending our localization theory to apply to degree sequences that follow a more relaxed property. Rather than require a strictly log-linear skew, we believe that a piece-wise log-linear degree sequence, or possibly even more general, will suffice to prove localization results nearly identical to our current theory. Such a result could  advance the literature to explain localization in a majority of real-world graphs. Our experiments also make it clear that the local clustering structure plays an important role in the behavior on real-world graphs.


\section*{Acknowledgments}
This research was supported by NSF CAREER award CCF-1149756 and DARPA SIMPLEX.

\bibliographystyle{plainnat}
\bibliography{newbib}

\begin{thebibliography}{41}
\providecommand{\natexlab}[1]{#1}
\providecommand{\url}[1]{\texttt{#1}}
\expandafter\ifx\csname urlstyle\endcsname\relax
  \providecommand{\doi}[1]{doi: #1}\else
  \providecommand{\doi}{doi: \begingroup \urlstyle{rm}\Url}\fi

\bibitem[Ahn et~al.(2012)Ahn, Guha, and McGregor]{Ahn:2012:GSS:2213556.2213560}
Kook~Jin Ahn, Sudipto Guha, and Andrew McGregor.
\newblock Graph sketches: Sparsification, spanners, and subgraphs.
\newblock In \emph{Proceedings of the 31st ACM SIGMOD-SIGACT-SIGAI Symposium on
  Principles of Database Systems}, PODS '12, pages 5--14, New York, NY, USA,
  2012. ACM.
\newblock ISBN 978-1-4503-1248-6.
\newblock \doi{10.1145/2213556.2213560}.
\newblock URL \url{http://doi.acm.org/10.1145/2213556.2213560}.

\bibitem[Andersen et~al.(2006)Andersen, Chung, and Lang]{andersen2006-local}
Reid Andersen, Fan Chung, and Kevin Lang.
\newblock Local graph partitioning using {PageRank} vectors.
\newblock In \emph{FOCS2006}, 2006.

\bibitem[Avrachenkov et~al.(2012)Avrachenkov, Litvak, Sokol, and
  Towsley]{avrachenkov2012quick}
Konstantin Avrachenkov, Nelly Litvak, Marina Sokol, and Don Towsley.
\newblock Quick detection of nodes with large degrees.
\newblock In Anthony Bonato and Jeannette Janssen, editors, \emph{Algorithms
  and Models for the Web Graph}, volume 7323 of \emph{Lecture Notes in Computer
  Science}, pages 54--65. Springer Berlin Heidelberg, 2012.
\newblock \doi{10.1007/978-3-642-30541-2_5}.

\bibitem[Bar-Yossef and Mashiach(2008)]{bar2008local}
Ziv Bar-Yossef and Li-Tal Mashiach.
\newblock Local approximation of pagerank and reverse pagerank.
\newblock In \emph{Proceedings of the 17th ACM conference on Information and
  knowledge management}, pages 279--288. ACM, 2008.

\bibitem[Bayati et~al.(2010)Bayati, Kim, and Saberi]{bayati2010graphgen}
Mohsen Bayati, JeongHan Kim, and Amin Saberi.
\newblock A sequential algorithm for generating random graphs.
\newblock \emph{Algorithmica}, 58\penalty0 (4):\penalty0 860--910, 2010.
\newblock ISSN 0178-4617.

\bibitem[Benzi and Razouk(2007)]{Benzi-2007-decay}
M~Benzi and N~Razouk.
\newblock Decay bounds and {O(n)} algorithms for approximating functions of
  sparse matrices.
\newblock \emph{ETNA}, 28:\penalty0 16--39, 2007.

\bibitem[Benzi and Golub(1999)]{benzi1999bounds}
Michele Benzi and Gene~H Golub.
\newblock Bounds for the entries of matrix functions with applications to
  preconditioning.
\newblock \emph{BIT Numerical Mathematics}, 39\penalty0 (3):\penalty0 417--438,
  1999.

\bibitem[Benzi et~al.(2013)Benzi, Boito, and Razouk]{Benzi-2013-decay}
Michele Benzi, Paola Boito, and Nader Razouk.
\newblock Decay properties of spectral projectors with applications to
  electronic structure.
\newblock \emph{SIAM Review}, 55\penalty0 (1):\penalty0 3--64, 2013.

\bibitem[Berkhin(2007)]{berkhin2007-bookmark}
Pavel Berkhin.
\newblock Bookmark-coloring algorithm for personalized {PageRank} computing.
\newblock \emph{Internet Mathematics}, 3\penalty0 (1):\penalty0 41--62, 2007.

\bibitem[Bonchi et~al.(2012)Bonchi, Esfandiar, Gleich, Greif, and
  Lakshmanan]{Bonchi-2012-fast-katz}
Francesco Bonchi, Pooya Esfandiar, David~F. Gleich, Chen Greif, and Laks~V.S.
  Lakshmanan.
\newblock Fast matrix computations for pairwise and columnwise commute times
  and {Katz} scores.
\newblock \emph{Internet Mathematics}, 8\penalty0 (1-2):\penalty0 73--112,
  2012.

\bibitem[Borgs et~al.(2014)Borgs, Brautbar, Chayes, and
  Teng]{borgs2014multiscale}
Christian Borgs, Michael Brautbar, Jennifer Chayes, and Shang-Hua Teng.
\newblock Multiscale matrix sampling and sublinear-time pagerank computation.
\newblock \emph{Internet Mathematics}, 10\penalty0 (1-2):\penalty0 20--48,
  2014.

\bibitem[Bressan and Pretto(2011)]{bressan2011local}
Marco Bressan and Luca Pretto.
\newblock Local computation of pagerank: the ranking side.
\newblock In \emph{Proceedings of the 20th ACM international conference on
  Information and knowledge management}, pages 631--640. ACM, 2011.

\bibitem[Bressan et~al.(2014)Bressan, Peserico, and
  Pretto]{bressan2014approximating}
Marco Bressan, Enoch Peserico, and Luca Pretto.
\newblock Approximating pagerank locally with sublinear query complexity.
\newblock \emph{arXiv preprint arXiv:1404.1864}, 2014.

\bibitem[Chung and Lu(2002)]{chung2002connected}
Fan Chung and Linyuan Lu.
\newblock Connected components in random graphs with given expected degree
  sequences.
\newblock \emph{Annals of combinatorics}, 6\penalty0 (2):\penalty0 125--145,
  2002.

\bibitem[Chung(1992)]{Chung-1992-book}
Fan R.~L. Chung.
\newblock \emph{Spectral Graph Theory}.
\newblock American Mathematical Society, 1992.

\bibitem[Clauset et~al.(2009)Clauset, Shalizi, and
  Newman]{clauset2009-powerlaw}
Aaron Clauset, Cosma~Rohilla Shalizi, and M.~E.~J. Newman.
\newblock Power-law distributions in empirical data.
\newblock \emph{SIAM Review}, 51\penalty0 (4):\penalty0 661--703, 2009.
\newblock \doi{10.1137/070710111}.
\newblock URL \url{http://link.aip.org/link/?SIR/51/661/1}.

\bibitem[Demko(1977)]{demko1977inverses}
Stephen Demko.
\newblock Inverses of band matrices and local convergence of spline
  projections.
\newblock \emph{SIAM Journal on Numerical Analysis}, 14\penalty0 (4):\penalty0
  616--619, 1977.

\bibitem[Demko et~al.(1984)Demko, Moss, and Smith]{demko1984decay}
Stephen Demko, William~F Moss, and Philip~W Smith.
\newblock Decay rates for inverses of band matrices.
\newblock \emph{Mathematics of computation}, 43\penalty0 (168):\penalty0
  491--499, 1984.

\bibitem[Faloutsos et~al.(1999)Faloutsos, Faloutsos, and
  Faloutsos]{faloutsos1999power}
Michalis Faloutsos, Petros Faloutsos, and Christos Faloutsos.
\newblock On power-law relationships of the internet topology.
\newblock In \emph{ACM SIGCOMM computer communication review}, 1999.

\bibitem[Fischler and Bolles(1981)]{fischler1981ransac}
Martin~A. Fischler and Robert~C. Bolles.
\newblock Random sample consensus: A paradigm for model fitting with
  applications to image analysis and automated cartography.
\newblock \emph{Commun. ACM}, 24\penalty0 (6):\penalty0 381--395, June 1981.
\newblock ISSN 0001-0782.
\newblock \doi{10.1145/358669.358692}.
\newblock URL \url{http://doi.acm.org/10.1145/358669.358692}.

\bibitem[Gleich(2010)]{bisquik}
David~F. Gleich.
\newblock {bisquik}: bisquik randomly samples a graph with a prescribed degree
  distribution.
\newblock \url{https://github.com/dgleich/bisquik}, November 2010.

\bibitem[Gleich(2015)]{Gleich-2015-prbeyond}
David~F. Gleich.
\newblock {PageRank} beyond the web.
\newblock \emph{SIAM Review}, 57\penalty0 (3):\penalty0 321--363, August 2015.
\newblock \doi{10.1137/140976649}.

\bibitem[Gleich and Kloster(2015)]{gleich2014sublinear}
David~F. Gleich and Kyle Kloster.
\newblock Sublinear column-wise actions of the matrix exponential on social
  networks.
\newblock \emph{Internet Mathematics}, 11\penalty0 (4-5):\penalty0 352--384,
  2015.

\bibitem[Gleich and Seshadhri(2012)]{Gleich-2012-neighborhoods}
David~F. Gleich and C.~Seshadhri.
\newblock Vertex neighborhoods, low conductance cuts, and good seeds for local
  community methods.
\newblock In \emph{KDD2012}, pages 597--605, August 2012.
\newblock \doi{10.1145/2339530.2339628}.

\bibitem[Higham(2008)]{Higham2008-functions-of-matrices}
Nicholas~J. Higham.
\newblock \emph{Functions of Matrices: Theory and Computation}.
\newblock SIAM, 2008.

\bibitem[Jeh and Widom(2003)]{jeh2003-personalized}
G.~Jeh and J.~Widom.
\newblock Scaling personalized web search.
\newblock In \emph{WWW}, pages 271--279, 2003.

\bibitem[Klimt and Yang(2004)]{klimt2004introducing}
Bryan Klimt and Yiming Yang.
\newblock Introducing the enron corpus.
\newblock In \emph{CEAS}, 2004.

\bibitem[Kolda et~al.(2014)Kolda, Pinar, Plantenga, and
  Seshadhri]{Kolda-2014-BTER}
Tamara~G. Kolda, Ali Pinar, Todd Plantenga, and C.~Seshadhri.
\newblock A scalable generative graph model with community structure.
\newblock \emph{SIAM Journal on Scientific Computing}, 36\penalty0
  (5):\penalty0 C424--C452, 2014.
\newblock \doi{10.1137/130914218}.

\bibitem[Leskovec and Sosi\v{c}(2014)]{snap}
Jure Leskovec and Rok Sosi\v{c}.
\newblock {SNAP}: A general purpose network analysis and graph mining library
  in {C++}.
\newblock \url{http://snap.stanford.edu/snap}, June 2014.

\bibitem[Leskovec et~al.(2005)Leskovec, Kleinberg, and
  Faloutsos]{leskovec2005shrinking-diameter}
Jure Leskovec, Jon Kleinberg, and Christos Faloutsos.
\newblock Graphs over time: Densification laws, shrinking diameters and
  possible explanations.
\newblock In \emph{Proceedings of the Eleventh ACM SIGKDD International
  Conference on Knowledge Discovery in Data Mining}, KDD '05, pages 177--187,
  New York, NY, USA, 2005. ACM.
\newblock ISBN 1-59593-135-X.
\newblock \doi{10.1145/1081870.1081893}.
\newblock URL \url{http://doi.acm.org/10.1145/1081870.1081893}.

\bibitem[Leskovec et~al.(2007)Leskovec, Adamic, and
  Huberman]{leskovec2007dynamics}
Jure Leskovec, Lada~A Adamic, and Bernardo~A Huberman.
\newblock The dynamics of viral marketing.
\newblock \emph{ACM Transactions on the Web (TWEB)}, 1\penalty0 (1):\penalty0
  5, 2007.

\bibitem[Leskovec et~al.(2009)Leskovec, Lang, Dasgupta, and
  Mahoney]{leskovec2009community}
Jure Leskovec, Kevin~J Lang, Anirban Dasgupta, and Michael~W Mahoney.
\newblock Community structure in large networks: Natural cluster sizes and the
  absence of large well-defined clusters.
\newblock \emph{Internet Mathematics}, 6\penalty0 (1):\penalty0 29--123, 2009.

\bibitem[Lofgren et~al.(2015)Lofgren, Banerjee, and
  Goel]{lofgren2015bidirectional}
Peter Lofgren, Siddhartha Banerjee, and Ashish Goel.
\newblock Bidirectional pagerank estimation: From average-case to worst-case.
\newblock In \emph{Algorithms and Models for the Web Graph}, pages 164--176.
  Springer, 2015.

\bibitem[Lofgren et~al.(2014)Lofgren, Banerjee, Goel, and
  Seshadhri]{lofgren2014fast}
Peter~A Lofgren, Siddhartha Banerjee, Ashish Goel, and C~Seshadhri.
\newblock Fast-ppr: Scaling personalized pagerank estimation for large graphs.
\newblock In \emph{Proceedings of the 20th ACM SIGKDD international conference
  on Knowledge discovery and data mining}, pages 1436--1445. ACM, 2014.

\bibitem[McSherry(2005)]{mcsherry2005-uniform}
Frank McSherry.
\newblock A uniform approach to accelerated {PageRank} computation.
\newblock In \emph{WWW}, pages 575--582, 2005.
\newblock ISBN 1-59593-046-9.

\bibitem[Mislove et~al.(2007)Mislove, Marcon, Gummadi, Druschel, and
  Bhattacharjee]{mislove-2007-socialnetworks}
Alan Mislove, Massimiliano Marcon, Krishna~P. Gummadi, Peter Druschel, and
  Bobby Bhattacharjee.
\newblock {Measurement and Analysis of Online Social Networks}.
\newblock In \emph{Proceedings of the 5th ACM/Usenix Internet Measurement
  Conference (IMC'07)}, San Diego, CA, October 2007.

\bibitem[Nassar et~al.(2015)Nassar, Kloster, and Gleich]{nassar2015strong}
Huda Nassar, Kyle Kloster, and David~F Gleich.
\newblock Strong localization in personalized pagerank vectors.
\newblock In \emph{Algorithms and Models for the Web Graph}, pages 190--202.
  Springer International Publishing, 2015.

\bibitem[Page et~al.(1999)Page, Brin, Motwani, and Winograd]{page1999-pagerank}
Lawrence Page, Sergey Brin, Rajeev Motwani, and Terry Winograd.
\newblock The {PageRank} citation ranking: Bringing order to the web.
\newblock Technical Report 1999-66, Stanford University, 1999.

\bibitem[Ripeanu et~al.(2002)Ripeanu, Foster, and
  Iamnitchi]{ripeanu2002mapping}
Matei Ripeanu, Ian Foster, and Adriana Iamnitchi.
\newblock Mapping the gnutella network: Properties of large-scale peer-to-peer
  systems and implications for system design.
\newblock \emph{arXiv preprint cs/0209028}, 2002.

\bibitem[Yang and Leskovec(2015)]{yang2015defining}
Jaewon Yang and Jure Leskovec.
\newblock Defining and evaluating network communities based on ground-truth.
\newblock \emph{Knowledge and Information Systems}, 42\penalty0 (1):\penalty0
  181--213, 2015.

\bibitem[Yin et~al.(2017)Yin, Benson, Leskovec, and
  Gleich]{Yin-2017-local-motif}
Hao Yin, Austin~R. Benson, Jure Leskovec, and David~F. Gleich.
\newblock Local higher-order graph clustering.
\newblock Accepted at KDD2017, 2017.

\end{thebibliography}
\normalsize
\appendix
\section{Appendix}\label{sec:app}

\subsection{Constructing seeded PageRank on complete-bipartite graphs}\label{sec:complete-bipartite}

To prove our de-localization results for seeded PageRank vectors on complete-bipartite graphs (Section~\ref{sec:negative}) we use an explicit construction of these vectors. We present that construction in this section.
Before going through the details of the construction, we first briefly outline the steps.

First we use the known structure of the spectrum of a complete-bipartite graph to construct a quadratic polynomial $q(x)$ that interpolates the scaled resolvent function $R(x) = (1-\alpha)\cdot(1- \alpha x)\inv$ on the spectrum of any complete-bipartite graph.
Then we study the block structure of the random-walk transition matrix $\mP$ for an arbitrary complete-bipartite graph.
Finally, we use this work to explicitly compute $q(\mP)\ve_j = R(\mP)\ve_j$ for any seed node $j$.

\paragraph{The construction}
We begin by noting that the random-walk transition matrix $\mP = \mA\mD^{-1}$ of an arbitrary complete-bipartite graph is necessarily diagonalizable. This is because it is a similarity transformation of a symmetric matrix: $\mD^{-1/2}\mP \mD^{1/2} = \mD^{-1/2}\mA\mD^{-1/2}$, where $\mA$ is symmetric because complete-bipartite graphs are undirected. Consequently we can use the following simple result from the theory of functions of matrices (\cite{Higham2008-functions-of-matrices}, Chapter 1.2): for any function $f(x)$ defined on the spectrum of $\mP$, we can express $f(\mP) = q(\mP)$ for any polynomial $q(x)$ that interpolates the function $f(x)$ on the spectrum of $\mP$.
If $\mP$ has $r$ distinct eigenvalues (i.e. $\mP$ has a minimal polynomial of degree $r$), then this implies that $f(\mP)$ can be expressed using an interpolating polynomial on $r$ values, and so $q(x)$ can be a polynomial of degree $r-1$.

Using the above facts, any matrix function $f(\mP)$ (such that $f(x)$ is defined on the spectrum of $\mP$) can be expressed in the form
\[
f(\mP) = c_0\mI + c_1\mP + \cdots + c_{r-1}\mP^{r-1}
\]
for appropriate coefficients $c_j$.
Next we show how to combine this fact with structural properties of complete-bipartite graphs to produce our de-localization results.
Note that we this addresses a broader set of functions than just the resolvent function, but later on we will study specifically the case of the resolvent function to obtain a more explicit bound on the de-localization of PageRank.

\paragraph{Specializing to complete-bipartite graphs}
The random-walk transition matrix $\mP$ of any complete-bipartite graph has eigenvalues $-1,0,$ and $1$ (\cite{Chung-1992-book}, Chapter 1.2). As noted above, $\mP$ is diagonalizable, so we can determine $f(\mP)$ by the action of $f(x)$ on the spectrum of $\mP$.
Thus, for any function $f(x)$ that is defined on the values $\{-1, 0, 1 \}$, and any polynomial $q(x)$ that satisfies
\begin{equation}\label{eqn:interpolating-polynomial}
q(-1) = f(-1), \quad q(0) = f(0), \quad p(1) = f(1)
\end{equation}
the polynomial will also satisfy $q(\mP) = f(\mP)$.
Below we give the structure of the interpolating polynomial for the resolvent function before showing how such polynomials can be used to prove that these functions are not local on complete-bipartite graphs.

The intuition of our results is that, because complete-bipartite graphs have 3 eigenvalues, any function of the graph can be expressed using an interpolating polynomial of degree 2; the matrices $\mP$ and $\mP^2$ then determine the structure of any function $f(\mP)$, and because of the uniform nature of complete-bipartite graphs, the expressions $\mP$ and $\mP^2$ (and, hence, $f(\mP)$) are roughly the sum of uniform distributions and so cannot be approximated using only a small number of entries that are non-zero.

\subsubsection{Interpolating polynomials}

Given any function $f(x)$ defined on the values $\{-1,0,1\}$, the degree 2 interpolating polynomial $q(x)$ for this function, defined by Equation~\eqref{eqn:interpolating-polynomial}, has coefficients $q(x) = c_0 + c_1 x + c_2 x^2$ given by
\begin{align}
c_0 &= f(0) \label{eqn:interpolating-coefficients0} \\
c_1 &= \tfrac{1}{2} ( f(1) - f(-1) ) \label{eqn:interpolating-coefficients1} \\
c_2 &= \tfrac{1}{2} ( f(1) + f(-1) - 2f(0) ). \label{eqn:interpolating-coefficients2}
\end{align}
The value $c_0 = f(0)$ follows from plugging 0 into $q(x)$ and noting $f(0) = q(0) = c_0 + 0 + 0 $.
The other values follow from similar straight-forward algebra.
Thus, for a general function $f(x)$ defined on the spectrum of a complete-bipartite graph, we can study the localization of
\begin{equation}\label{eqn:general-function-bipartite}
f(\mP) = f(0) \cdot \mI + \left( \tfrac{1}{2} ( f(1) - f(-1) ) \right) \cdot \mP + \left( \tfrac{1}{2} ( f(1) + f(-1) - 2f(0) ) \right) \cdot \mP^2,
\end{equation}
by looking at the structure of the matrices $\mP$ and $\mP^2$. Because we intend to focus on the particular choice of the resolvent function, $f(x) = (1-\alpha)\cdot (1- \alpha x)\inv$, we compute
\begin{align}
c_0 &= f(0) = (1-\alpha) \\
c_1 &= \tfrac{1}{2} ( f(1) - f(-1) ) = \tfrac{\alpha}{1+\alpha} \\
c_2 &= \tfrac{1}{2} ( f(1) + f(-1) - 2f(0) ) = \tfrac{\alpha^2}{1+\alpha}
\end{align}
so that we can write the more specific expression
\[
f(\mP)\ve_j = (1-\alpha) \ve_j + \tfrac{\alpha}{1+\alpha}\mP\ve_j + \tfrac{\alpha^2}{1+\alpha}\mP^2 \ve_j
\]
for a seeded PageRank vector on a complete-bipartite graph. All that remains is to analyze the block structure of $\mP$ itself to determine the exact de-localization behavior of PageRank in this setting.

\subsubsection{Using graph structure}
\label{sec:using-graph-structure}
Here we will consider a complete-bipartite graph with $n$ nodes divided into a partition of $k$ nodes and $n-k$ nodes. In particular we want to understand the structure of the random-walk transition matrix, $\mP$, as well as $\mP^2$, for such graphs.
The nodes can be ordered so that the adjacency matrix has the block form
$\mA = \bmat{0 & \mE^T \\ \mE & 0}$, where $\mE$ is the $(n-k)\times k$ matrix of all 1s. The degree matrix is then $\mD = \bmat{(n-k)\cdot \mI_{n-k} & 0\\ 0& k \cdot \mI_k}$.
Letting $\bmat{\mI_k \\ 0}$ be a $n\times k$ matrix with a $k\times k$ identity block in the upper-most block, we can express the relationships
\begin{align}
\mP &= \mA\mD\inv = \tfrac{1}{k}\bmat{\mI_k \\ 0}\bmat{0 & \mE^T} + \tfrac{1}{n-k}\bmat{0\\ \mE }\bmat{ \mI_k & 0} \\
&= \tfrac{1}{k}\bmat{0&\mE^T\\0&0} + \tfrac{1}{n-k}\bmat{0&0\\ \mE &0} \label{eqn:bipartite-P}
\end{align}
and
\begin{align}
\mP^2 &=  \tfrac{1}{n-k}\tfrac{1}{k} \bmat{0&0\\0& \mE\mE^T} + \tfrac{1}{n-k}\tfrac{1}{k} \bmat{\mI_k\\0} \mE^T\mE\bmat{\mI_k & 0} \\
&= \tfrac{1}{n-k}\bmat{0&0\\0& \mJ_{n-k} } + \tfrac{1}{k}\bmat{\mJ_k &0\\0&0} \label{eqn:bipartite-P2}
\end{align}
where $\mJ_i$ is a square block of 1s of dimension $i\times i$. We use the notation $\mJ_i$ specifically to denote square matrices of all 1s, whereas we use $\mE$ to denote matrices of all 1s that are possibly not square.
Equation~\eqref{eqn:bipartite-P2} uses the relationships $\mE\mE^T = k\cdot \mJ_{n-k}$ and $\mE^T\mE = (n-k)\cdot \mJ_k$, which follow from the fact that the matrix $\mE$ is all 1s and dimension $(n-k)\times k$.

Next we use the expressions for $\mP$ and $\mP^2$ in Equations~\eqref{eqn:bipartite-P} and \eqref{eqn:bipartite-P2}
to show that columns of matrix functions
$f(\mP) \ve_j$ are de-localized, except for extreme cases which we discuss below. To complete our proof,
we will simply substitute the above expressions for $\mP^t$ in Equations~\eqref{eqn:bipartite-P} and \eqref{eqn:bipartite-P2}
into Equation~\eqref{eqn:general-function-bipartite} and perform a little algebra.
First consider any node $j$ in the partition containing $k$ nodes. From the structure of $\mP$ described above, we get that $\mP\ve_j = \tfrac{1}{n-k} \bmat{ 0 \\ \ve }_{n-k}$, where $\bmat{ 0 \\ \ve }_{n-k}$ is a vector with $k$ entries that are $0$s, and $(n-k)$ entries that are 1s in the lower block. Similarly we obtain that $\mP^2\ve_j = \tfrac{1}{k} \bmat{ \ve \\ 0 }_k$, where $\bmat{ \ve \\ 0 }_k$ is a vector with $k$ entries that are 1s in the upper block, followed by $(n-k)$ zero entries.

Now we use this information to compute the magnitude of the individual entries of the vectors $f(\mP)\ve_j$. Substituting into Equation~\eqref{eqn:general-function-bipartite}
 the structure of $\mP \ve_j$ and $\mP^2 \ve_j$ that we just computed, we can see that $f(\mP)\ve_j$ has $k$ entries of magnitude exactly $\tfrac{1}{k} c_2 $, and $n-k-1$ entries of magnitude exactly $\tfrac{1}{n-k} c_1 $, and finally entry $j$ itself has magnitude $f(0) +  \tfrac{1}{n-k} c_1$.

The purpose of this arithmetic is that it allows us to see that a vector $f(\mP)\ve_j$ has too many non-trivial entries, and so
cannot be approximated to an accuracy of $\eps$ with fewer than $O(n)$ of its non-zero entries, unless the function $f(x)$ is such that
at least one of the coefficients $\tfrac{1}{n-k}c_1$ or $\tfrac{1}{k} c_2$
is very small relative to $\eps$ (where the coefficients $c_t$ are as defined above in Equation~\eqref{eqn:interpolating-coefficients0}).
We make this notion rigorous in the following proposition, which the above discussion has proved:

\begin{theorem}\label{thm:general-delocalized-bipartite}
Let $\mP$ be the random-walk transition matrix of a complete-bipartite graph on $n$ nodes, and let the partition sizes be $k$ and $n-k$. Fix a value $\alpha \in (0,1)$ and let $f(x)$ be the resolvent function $f(x) = (1-\alpha)\cdot(1-\alpha x)\inv$.
Then columns of $f(\mP)$ can be expressed as follows. For any node $j$ in the partition of size $k$, we have, in the notation of Section~\ref{sec:using-graph-structure},
\begin{equation}\label{eqn:explicit-pr-vector}
f(\mP)\ve_j = (1-\alpha) \cdot \ve_j + \tfrac{\alpha}{(1+\alpha)}\tfrac{1}{(n-k)} \cdot \bmat{ 0 \\ \ve }_{n-k}
+ \tfrac{\alpha^2}{(1+\alpha)}\tfrac{1}{k}\cdot \bmat{\ve \\ 0}_k.
\end{equation}

\end{theorem}

This result implies that every column of $R(\mP)$ consists of one large entry at the seed node, $R(\mP)_{jj}$, and one sub-vector for each of the two graph partitions, with each sub-vector having entries of uniform value.
Note that the sum of the entries in the partition containing the seed is $\tfrac{1}{1+\alpha}$, and the sum of the entries in the other partition is then $\tfrac{\alpha}{1+\alpha}$.

When studying the degree-normalized error it will be useful for us also to have explicit computation of the vector $\mD\inv\vx = \mD\inv R(\mP)\ve_j$. In complete-bipartite graphs, nodes in the partition of size $k$ each have degree $n-k$, and nodes in the partition of size $n-k$ have degree $k$. Thus, scaling the PageRank vector in~\eqref{eqn:explicit-pr-vector} by node degrees yields the following.
\begin{corollary}\label{thm:degree-normalized-cbg}
In the notation of Theorem~\ref{thm:general-delocalized-bipartite}, the degree-normalized seeded PageRank vector seeded on node $j$ is
\begin{equation}\label{eqn:explicit-pr-deg}
\mD\inv R(\mP)\ve_j = (1-\alpha) \tfrac{1}{n-k}\cdot \ve_j + \tfrac{\alpha}{(1+\alpha)}\tfrac{1}{k(n-k)} \cdot \bmat{ 0 \\ \ve }_{n-k}
+ \tfrac{\alpha^2}{(1+\alpha)}\tfrac{1}{k(n-k)}\cdot \bmat{\ve \\ 0}_k.
\end{equation}
\end{corollary}
Having stated these preliminary constructions, we next proceed with proofs of our localization and de-localization results.

\subsection{Proofs of de-localization results on complete-bipartite graphs}\label{sec:delocalization-results}

\subsubsection{Localization behavior in the 1-norm}\label{sec:delocalization-1norm}
Recall that the overarching goal of the paper is to study the sparsest possible approximation to a seeded PageRank vector, such that the approximation meets a desired accuracy criterion. In this section we use the explicit PageRank construction in Theorem~\ref{thm:general-delocalized-bipartite} to study the sparsity of approximations that are $\eps$-accurate in the 1-norm and degree-normalized 1-norm, where $\eps$ is the desired accuracy.

We construct our sparse approximation as follows. Let $\hvv{x}$ be the vector $(1-\alpha)\ve_j$ plus a certain number of the nonzero entries in the true solution $\vx$.
More specifically, $\hvv{x}$ equals $(1-\alpha)\ve_j$ plus $z_1$ entries of $\vx$ corresponding to nodes in the partition of size $n-k$, and $z_2$ entries of $\vx$ corresponding to nodes in the partition of size $k$. Note that we design $\hvv{x}$ including the component $(1-\alpha)\ve_j$ simply to make the error analysis cleaner; this only affects the sparsity of the approximation by 1 and makes no difference in the asymptotic results. Thus, the approximation we are constructing is within 1 nonzero entry of being the sparsest possible.

Next we want to measure the 1-norm and 2-norm error of the approximation $\hvv{x}$ that results from retaining $z_1 + z_2$ nonzeros as described above. We begin with the 1-norm analysis. We claim that the error satisfies
\begin{equation}\label{eqn:1norm-retaining-ineq}
\| \vx - \hvv{x} \|_1 = \tfrac{\alpha}{1+\alpha}\left(1 - \tfrac{z_1}{n-k}\right) + \tfrac{\alpha^2}{1+\alpha}\left(1- \tfrac{z_2}{k}\right).
\end{equation}
To see this, note that the entries in $\vx$ corresponding to nodes in the seed's partition sum to $\alpha^2/(1+\alpha)$, whereas nodes outside the seed's partition sum to $\alpha/(1+\alpha)$. If $z_1$ entries outside the seed's partition are retained, then $(n-k)- z_1$ entries are omitted from that sub-vector, so the error accrued from that sub-vector alone is $((n-k)- z_1)/(n-k) = (1 - z_1/(n-k) )$ times the total error contained in those nodes, which is $\alpha/(1+\alpha)$. Similar arithmetic yields the second summand of Equation~\eqref{eqn:1norm-retaining-ineq}.

To construct a \emph{sparse} approximation of PageRank we want to retain the \emph{largest} nonzeros in $\vx$. This requires that we know which entries have the largest and smallest value.
There are two cases: the case in which the nodes in the seed's partition are larger in magnitude, i.e.~ when $\tfrac{\alpha^2}{k(1+\alpha)} > \tfrac{\alpha}{(n-k)(1+\alpha)}$,
and the case in which they are not.
Simple algebraic manipulation shows that these cases are determined by the following inequality. Nodes in the partition of the seed (the partition of size $k$) have larger PageRank value if and only if $k < \alpha(n-k)$.

First we consider that case that the seed's partition has larger PageRank values, i.e.~$k < \alpha(n-k)$. If this is the case, we want to retain nodes in the seed's partition (because they are larger in magnitude). In the context of Equation~\eqref{eqn:1norm-retaining-ineq} this means we are increasing $z_2$. However, even if we retain all nodes in the seed's partition (by setting $z_2 = k$), the total 1-norm error can still be as large as $\alpha/(1+\alpha)$ if we do not retain any nonzeros in the other partition (this corresponds to setting $z_1 = 0$).
If $\eps < \alpha/(1+\alpha)$, then we must increase $z_1$ in order to obtain an $\eps$-accurate approximation.

We remark that if $\eps < \alpha/(1+\alpha)$, then we \emph{must} retain $z_2 = k$ nonzeros just to attain an accuracy of $\alpha/(1+\alpha)$, and if $k = \Theta(n)$, then our approximation is already de-localized. However, to cover the case that $k$ is small, we continue our analysis.

If we retain all $k$ nodes in the seed's partition, and only $z_1$ nodes in the other partition, then by Equation~\eqref{eqn:1norm-retaining-ineq} the 1-norm error is $\alpha/(1+\alpha)(1 - z_1/(n-k))$. This is bounded above by $\eps$ if and only if
\[
\left( 1 - \eps \tfrac{1+\alpha}{\alpha}\right)(n-k) \leq z_1.
\]
The total number of nonzeros in the approximation $\hvv{x}$ would then be $z_1 + z_2$, which is
\begin{align}
z_1+k &\geq \left( 1 - \eps \tfrac{1+\alpha}{\alpha}\right)(n-k) + k \\
&=
\left( 1 - \eps \tfrac{1+\alpha}{\alpha}\right)n + \left(\eps \tfrac{1+\alpha}{\alpha}\right) k,
\end{align}
which completes a proof in the first case (the case $k < \alpha(n-k)$) that $\Theta(n)$ nonzeros are required for an $\eps$-accurate approximation.

In the second case, $k \geq \alpha(n-k)$, nodes outside the seed's partition are greater in magnitude, and so we retain those first. Similar to the previous case, setting $z_1 = (n-k)$ and $z_2=0$ in Equation~\eqref{eqn:1norm-retaining-ineq} yields an approximation with a 1-norm accuracy of $\tfrac{\alpha^2}{1+\alpha}$. Again, $(n-k)$ might already be enough nonzeros to be de-localized, but we continue our analysis for the case that $(n-k)$ is small.

If $\tfrac{\alpha^2}{(1+\alpha)} > \eps$ then additional nonzeros must be retained in our approximation. Hence, we must increase $z_2$. Note that the 1-norm error of our approximation is $\tfrac{\alpha^2}{(1+\alpha)}(1 - z_2/k)$. This is bounded above by $\eps$ if and only if
\[
\left( 1 - \eps \tfrac{1+\alpha}{\alpha^2}\right)k \leq z_2.
\]
The total number of nonzeros in the approximation $\hvv{x}$ is, again, $z_1 + z_2$, which is
\begin{align}
n-k+z_2 &\geq (n-k) + \left( 1 - \eps \tfrac{1+\alpha}{\alpha^2}\right)k \\
&= n - \left(\eps \tfrac{1+\alpha}{\alpha^2}\right) k.
\end{align}
Since $k \leq n$ we have $-k \geq - n$, and so we can write $n-k+z_2 \geq (1 - \eps(1+\alpha)/\alpha^2)n$
which completes the second case, $k \geq \alpha(n-k)$.
Finally, both cases guarantee that approximating seeded PageRank with a 1-norm accuracy of $\eps$ requires at least $(1 - \eps (1+\alpha)/\alpha^2) n$ nonzero entries, proving the following result.

\def\thetheorem{\ref{thm:localized-1norm}}
\begin{proposition*}
Let $\mP$ be the random-walk transition matrix of an $n$-node complete-bipartite graph, and let $j$ be the index of any node in the graph.
Fix a value $\alpha \in (0,1)$, and a desired accuracy $\eps < \alpha^2/(1+\alpha)$.
Then the number of nonzeros required to approximate the seeded PageRank vector $(1-\alpha)(\mI-\alpha\mP)\inv\ve_j$ with a 1-norm accuracy of $\eps$ is bounded below by
$(1 - \eps (1+\alpha)/\alpha^2)n$.
\end{proposition*}
This statement implies that any seeded PageRank vector in any complete-bipartite graph is de-localized when error is measured in the 1-norm.
The bound given in the proposition is not tight -- rather, we relax the bounds that we gave in our proofs above for the sake of a cleaner, more intuitive statement.

\paragraph{Degree-normalized 1-norm}
Next we consider localization behavior when we normalize error by node degrees.
More specifically, here we study the localization when we measure error using the expression $\| \mD\inv( \vx - \hvv{x}) \|_1$.

This analysis builds on our previous analysis simply by scaling the nodes' PageRank values by the node degrees. Recalling Corollary~\ref{thm:degree-normalized-cbg}, we will analyze the error in approximating the following scaled PageRank vector:
\begin{equation}\label{eqn:deg-explicit-pr-vector}
\mD\inv R(\mP)\ve_j = (1-\alpha) \tfrac{1}{n-k}\cdot \ve_j + \tfrac{\alpha}{(1+\alpha)}\tfrac{1}{k(n-k)} \cdot \bmat{ 0 \\ \ve }_{n-k}
+ \tfrac{\alpha^2}{(1+\alpha)}\tfrac{1}{k(n-k)}\cdot \bmat{\ve \\ 0}_k.
\end{equation}

First we show that the approximation $\hvv{x} = (1-\alpha)\ve_j$  can be $\eps$-accurate, depending on the sizes of $n$ and the partitions of the complete-bipartite graph. This will give an example of a sparse approximation of PageRank that is $\eps$-accurate, showing that, when error is measured as $\| \mD\inv(\cdot)\|_1$,
seeded PageRank can be localized in the context of complete-bipartite graphs.

Observe from Equation~\eqref{eqn:deg-explicit-pr-vector} that the error of this sparse approximation is
\begin{align*}
\left\| \mD\inv\left(\vx - (1-\alpha)\ve_j\right)\right\|_1 &= \tfrac{\alpha}{(1+\alpha)}\tfrac{1}{k}
+ \tfrac{\alpha^2}{(1+\alpha)}\tfrac{1}{(n-k)} \\
&= \tfrac{\alpha}{(1+\alpha)} \left(\tfrac{1}{k}
+ \tfrac{\alpha}{(n-k)}\right).
\end{align*}
For a complete-bipartite graph with balanced partitions (partitions with sizes $k = n/2 = (n-k)$) the error of this approximation is $(\alpha/(1+\alpha))(1+\alpha)2/n = 2\alpha/n$. As $n$ increases this quantity goes to 0, proving that, for large enough graphs, the sparse approximation $\hvv{x} = (1-\alpha)\ve_j$  has error less than $\eps$ when measured in the degree-normalized 1-norm.
This result demonstrates that seeded PageRank can be localized on dense graphs (since a complete bipartite graph with partitions of size $n/2$ has $O(n^2)$ edges) when error is measured in the degree-normalized 1-norm.

\textbf{Sparse $K_{k,n-k}$.}
In contrast with the above example of a dense graph, this paper is primarily concerned with \emph{sparse} graphs -- graphs with $O(n)$ edges. This is because real-world graphs that exhibit localization behavior tend to be sparse.
Hence, next we consider specifically complete-bipartite graphs in which at least one partition is of a constant size, as this class of graph has only $O(n)$ edges.

Forcing one partition to have constant size creates two cases: the case in which the seed is in the partition of constant size ($k = O(1)$) and the case in which it is not ($(n-k) = O(1)$).
Next we show in each of these two cases that seeded PageRank vectors are de-localized (in the context of sparse complete-bipartite graphs) when error is measured in the degree-normalized 1-norm.

First we address the case that the seed is in the small partition, so $k = O(1)$. In this case, our approximation $\hvv{x}$ can retain all $k$ nonzeros in $\vx$ that correspond to nodes in the small partition, and the approximation will still be localized. This leaves $n-k$ entries in $\vx$ that we have not yet retained, all of magnitude $\alpha/( k(n-k)(1+\alpha) )$. If our approximation $\hvv{x}$ retains $z$ of these nonzeros (in addition to the other $k$ nonzeros already retained), then the error is
\[
\|\mD\inv(\vx - \hvv{x})\|_1 = \tfrac{\alpha}{ (1+\alpha)}\tfrac{1}{k}\left(1 - \tfrac{z}{n-k}\right).
\]
Thus, in order to guarantee $\|\mD\inv(\vx - \hvv{x})\|_1 \leq \eps$, the number of nonzeros retained must satisfy
\[
\left(1- \eps \tfrac{(1+\alpha)k}{\alpha}\right) (n-k) \leq z.
\]
Because we assumed that $k = O(1)$, we know $(n-k) = O(n)$, and so this proves that the PageRank vector $\vx$ is de-localized, as long as  $\eps < \tfrac{\alpha}{k (1+\alpha)}$. (We note that this is a constant because $k$ is assumed to be a constant.)

Next we address the case that the seed is in the large partition, which is the case $k = \Theta(n)$. In this case our approximation can retain all $n-k = O(1)$ nonzeros in the small partition and the approximation will still be sparse. This leaves $k-1$ entries of magnitude $\alpha^2/( k(n-k)(1+\alpha) )$, and the seed node. To make the analysis a little cleaner, let our approximation be $\hvv{x} = (1-\alpha)\ve_j$ plus the $n-k$ nonzeros in the small partition of the graph.
If, in addition, our approximation retains $z$ of the $k$ remaining nonzeros, then the error is
\[
\|\mD\inv(\vx - \hvv{x})\|_1 = \tfrac{\alpha^2}{ (1+\alpha)}\tfrac{1}{n-k}\left(1 - \tfrac{z}{k}\right).
\]
Thus, in order to guarantee $\|\mD\inv(\vx - \hvv{x})\|_1 \leq \eps$, the number of nonzeros retained must satisfy
\[
\left(1- \eps \tfrac{(1+\alpha)(n-k)}{\alpha^2}\right) k \leq z.
\]
Because we assumed that $k = \Theta(n)$, we know $(n-k) = O(1)$, and so this proves that the PageRank vector $\vx$ is de-localized, as long as  $\eps < \tfrac{\alpha }{(1+\alpha)(n-k)}$. (We note again that this is a constant because here we assume $n-k$ is a constant.)
This completes our proof that seeded PageRank vectors on sparse complete-bipartite graphs are de-localized in the degree-normalized 1-norm.

\subsubsection{Localization behavior in the 2-norm}\label{sec:local-2norm}

To study how localization behaves when error is measured in the 2-norm, we return to Equation~\eqref{eqn:explicit-pr-vector}. The analysis we performed above (to relate sparsity of the approximation to $\alpha$ and $\eps$) carries over to the 2-norm, with only mild changes.

The big difference (that seeded PageRank on complete-bipartite graphs is \emph{always} localized when error is measured in the 2-norm) is a consequence of the following relationship of the 1-norm and 2-norm. The best example of a de-localized vector is the uniform distribution, the length $n$ vector $(1/n)\ve$. Note that the 1-norm of this vector is 1, but the 2-norm is $1/\sqrt{n}$. Thus, as $n$ increases, this uniform distribution stays de-localized in the 1-norm, but can be ``accurately approximated" with the 0 vector if error is measured in the 2-norm: $\| (1/n)\ve - 0 \|_2 = 1/\sqrt{n}$, which is bounded by  $\eps$ as $n$ increases.
This same effect makes seeded PageRank vectors ``localized" when error is measured in the 2-norm, because, in our setting of complete-bipartite graphs, the PageRank vectors are essentially two uniform distributions glued together (see the equation in Theorem~\ref{thm:general-delocalized-bipartite}).

Here we demonstrate this localization behavior more rigorously. Following the analysis in Section~\ref{sec:delocalization-1norm}, consider the squared 2-norm error resulting from the approximation $\hvv{x} = (1-\alpha)\ve_j$ plus $z_1$ non-zero entries retained outside the seed's partition, and $z_2$ nonzeros retained in the seed's partition. To compute the squared 2-norm error $\| \vx - \hvv{x}\|_2^2$, we must square all entries that we do not retain and then sum the result. This produces the following modification of Equation~\eqref{eqn:1norm-retaining-ineq} which we derived when measuring error in the 1-norm:
\begin{equation}\label{eqn:2norm-retaining-ineq}
\| \vx - \hvv{x} \|_2^2 = \left(\tfrac{\alpha}{1+\alpha}\right)^2\tfrac{1}{(n-k)}\left(1 - \tfrac{z_1}{n-k}\right) + \left(\tfrac{\alpha^2}{1+\alpha}\right)^2\tfrac{1}{k}\left(1- \tfrac{z_2}{k}\right).
\end{equation}
The factor of $1/(n-k)$ comes from the fact that in $\vx$ there are $(n-k)$ entries of magnitude $(\alpha/(1+\alpha))(1/(n-k))$ that must be squared and summed to compute the 2-norm. (The factor of $1/k$ has a similar origin.)

Equation~\eqref{eqn:2norm-retaining-ineq} allows us to show that seeded PageRank can always be sparsely approximated with a 2-norm accuracy of $\eps$. If both $(n-k)$ and $k$ are large, then we can set $z_1 = 0$ and $z_2 = 0$ and obtain an $\eps$ accurate approximation; in other words, $\hvv{x} = (1-\alpha)\ve_j$ would be an $\eps$-accurate approximation in the 2-norm. This is because  setting $z_1,z_2= 0$ results in the error
\begin{align*}
\| \vx - (1-\alpha)\ve_j \|_2^2 &= \left(\tfrac{\alpha}{1+\alpha}\right)^2\tfrac{1}{(n-k)} + \left(\tfrac{\alpha^2}{1+\alpha}\right)^2\tfrac{1}{k}
\end{align*}
which goes to 0 if both $k$ and $(n-k)$ increase.

In the case that one partition stays constant size, then we set $\hvv{x}$ to retain all nonzeros in the partition of constant size. If $k = O(1)$, then this results in an approximation $\hvv{x}$ with no more than $k+1$ nonzeros, and an error of $(\alpha/(1+\alpha))^2/(n-k)$, which goes to zero as $n$ increases (since we have assumed $k$ is bounded by a constant). On the other hand, if $(n-k) = O(1)$, then retaining all $(n-k)$ nonzeros still results in an approximation with a constant number of nonzeros, and the error will be $(\alpha^2/(1+\alpha))^2/k$. Again, this goes to 0 as $n$ increases, because $(n-k) = O(1)$ implies $k = (n)$, and so $k$ would go to infinity as well.

This completes a proof that seeded PageRank on complete-bipartite graphs can be approximated with a 2-norm accuracy of $\eps$ using only a constant number of nonzeros.

\textbf{Degree-normalized 2-norm.}
Here we see the same localization behavior for seeded PageRank vectors on complete-bipartite graphs when measuring error in a degree-normalized 2-norm. The proof follows the proofs above with only minor modifications to account for the degree scaling.

From Equation~\eqref{eqn:explicit-pr-deg} we have
\begin{equation}\label{eqn:2norm-deg-retaining-ineq}
\| \mD\inv(\vx - \hvv{x}) \|_2^2 = \left(\tfrac{\alpha}{1+\alpha}\right)^2\tfrac{1}{k^2(n-k)}\left(1 - \tfrac{z_1}{n-k}\right) + \left(\tfrac{\alpha^2}{1+\alpha}\right)^2\tfrac{1}{(n-k)^2k}\left(1- \tfrac{z_2}{k}\right),
\end{equation}
where again $z_1$ and $z_2$ are the number of nonzeros from the two graph partitions that we retain in the approximate solution.
Setting $z_1,z_2 = 0$, (once again using the sparse approximation $\hvv{x} = (1-\alpha)\ve_j$) we get the squared, degree-normalized 2-norm error
\begin{align*}
\| \mD\inv(\vx - \hvv{x}) \|_2^2 &= \tfrac{1}{k(n-k)}\left(
\left(\tfrac{\alpha}{1+\alpha}\right)^2\tfrac{1}{k} + \left(\tfrac{\alpha^2}{1+\alpha}\right)^2\tfrac{1}{(n-k)}
\right) \\
&\leq \tfrac{1}{k(n-k)},
\end{align*}
which is bounded by $1/n$.
Hence, we have that $\| \mD\inv(\vx - \hvv{x})\|_2 \leq 1/\sqrt{n}$, which goes to 0 as $n$ increases. This proves that seeded PageRank can be accurately approximated with a constant number of non-zeros when error is measured in the degree-normalized 2-norm.

\subsubsection{Localization behavior in residual vectors}\label{sec:local-residuals}

As a final observation on the relationship of norms to de-localization behavior, we look at the localization of \emph{residual} vectors related to approximate PageRank vectors. Many algorithms for rapidly approximating seeded PageRank vectors in a sparse manner (such as the Gauss-Southwell linear solver we use in this paper) operate by maintaining sparse iterative solution $\hvv{x}$ and residual vectors $\vr = (1-\alpha)\ve_j - (\mI-\alpha\mP)\hvv{x}$.
If convergence is desired in the 1-norm then often the exact error can be computed simply by studying $\|\hvv{x}\|_1$. This is possible if $\vx \geq \hvv{x}\geq 0$ holds entry-wise. However, if convergence is desired in other norms (for example, $\|\mD\inv(\cdot)\|_{\infty}$ is used in~\cite{andersen2006-local}) then the residual is used to determine convergence via the relationship $\vx - \hvv{x} = (1-\alpha)(\mI-\alpha\mP)\inv\vr$.

Here we construct a toy example to show that, depending on which norm is used, it is possible to have a sparse residual with norm less than $\eps$, even if the exact solution is de-localized and the approximate solution has poor accuracy. The purpose of such an example is to show that we must carefully choose the norms we used to determine convergence of our algorithms for seeded PageRank, especially when the residual is used to determine convergence.

The toy example is simply the seeded PageRank vector for the star graph on $n$ nodes, where the seed is the center node, which we will index as node 1. In this setting, the trivial approximation $\hvv{x} = 0$ has residual $\vr = (1-\alpha)\ve_1$. Using the degree-normalized 1- or 2-norm suggests that the approximation is good:
\begin{align*}
\| \mD\inv \vr \|_1 &= \| \mD\inv (1-\alpha) \ve_1\|_1
= \tfrac{(1-\alpha)}{(n-1)} \\
\| \mD\inv \vr \|_2 &= \| \mD\inv (1-\alpha) \ve_1\|_2
= \tfrac{(1-\alpha)}{(n-1)} .
\end{align*}
In both norms, the residual has norm $(1-\alpha)/(n-1)$, which goes to 0 as $n$ increases.
This is problematic because the actual solution vector, computed via Theorem~\ref{thm:general-delocalized-bipartite}, is $\vx = (1/(1+\alpha))\ve_1 + (\alpha/(1+\alpha))/(n-1)(\ve-\ve_1)$ -- as a uniform distribution with one altered entry, this vector is totally de-localized, yet a sparse, inaccurate approximation ($\hvv{x} = 0$) yields a sparse residual ($(1-\alpha) \ve_1$) with seemingly high accuracy, $(1-\alpha)/(n-1)$, when viewed through the degree-normalized 1- and 2-norms. In contrast, both the 1- and 2-norms report a large norm for this residual: $\| (1-\alpha) \ve_1 \|_1 = \|(1-\alpha) \ve_1\|_2 = (1-\alpha)$. Again, this suggests that non-degree-normalized norms are more trustworthy than degree-normalized norms for detecting localization behavior in seeded PageRank vectors.

\end{document}